\documentclass[sigconf]{acmart}
\usepackage{xspace}
\usepackage{enumitem}
\usepackage{caption}
\usepackage{multirow}
\usepackage{subcaption}
\usepackage{booktabs}

\newcommand{\name}{PrivacyAkinator\xspace}

\newcommand{\revise}[1]{\textcolor{blue}{#1}}
\newcommand{\sssec}[1]{\vspace*{0.05in}\noindent\textbf{#1}}

\AtBeginDocument{%
  }

\copyrightyear{2026}
\acmYear{2026}
\setcopyright{cc}
\setcctype{by}
\acmConference[CHI '26]{Proceedings of the 2026 CHI Conference on Human Factors in Computing Systems}{April 13--17, 2026}{Barcelona, Spain}
\acmBooktitle{Proceedings of the 2026 CHI Conference on Human Factors in Computing Systems (CHI '26), April 13--17, 2026, Barcelona, Spain}
\acmPrice{}
\acmDOI{10.1145/3772318.3790408}
\acmISBN{979-8-4007-2278-3/2026/04}

\begin{document}

\title{\name: Articulating Key Privacy Design Decisions by Answering LLM-Generated Multiple-choice Questions}

\author{Qiyu Li}
\affiliation{%
  \institution{University of California San Diego}
  \city{La Jolla, California}
  \country{USA}}
\email{qiyuli@ucsd.edu}

\author{Yuen Sum Wong}
\affiliation{%
  \institution{University of California San Diego}
  \city{La Jolla, California}
  \country{USA}}
\email{y6wong@ucsd.edu}

\author{Yuen Kei Wong}
\affiliation{%
  \institution{University of California San Diego}
  \city{La Jolla, California}
  \country{USA}}
\email{ykw001@ucsd.edu}

\author{Longxuan Yu}
\affiliation{%
  \institution{University of California Riverside}
  \city{Riverside, California}
  \country{USA}}
\email{ylong030@ucr.edu}

\author{Haojian Jin}
\affiliation{%
  \institution{University of California Riverside}
  \city{La Jolla, California}
  \country{USA}}
\email{haojian@ucsd.edu}

\renewcommand{\shortauthors}{Li et al.}


%

\begin{abstract}
NIST's Privacy Risk Assessment Methodology (PRAM) provides a structured framework for privacy experts to assess privacy risks. However, its complexity and reliance on expert knowledge make it difficult for novice developers to use effectively. This paper explores methods to lower these barriers. We first performed an observational study with 12 participants using PRAM in real-world scenarios, and found that novice developers struggled most with articulating privacy-related design decisions. We then developed \name, an interactive tool that helps developers articulate key privacy decisions by answering LLM-generated multiple-choice questions. \name introduces three innovations: a universal privacy representation that abstracts privacy-related design decisions into data flows and stakeholder interactions; a domain-aware design space mined from 10K privacy-related news articles; and a dynamic question-generation workflow to prioritize relevant questions. 
Our user study with 24 participants suggests that developers using \name identified 47\% more key decisions in 73\% less time compared to PRAM.

\end{abstract}











\begin{teaserfigure}
    \makebox[\textwidth][c]{
    \includegraphics[width=1.05\textwidth]{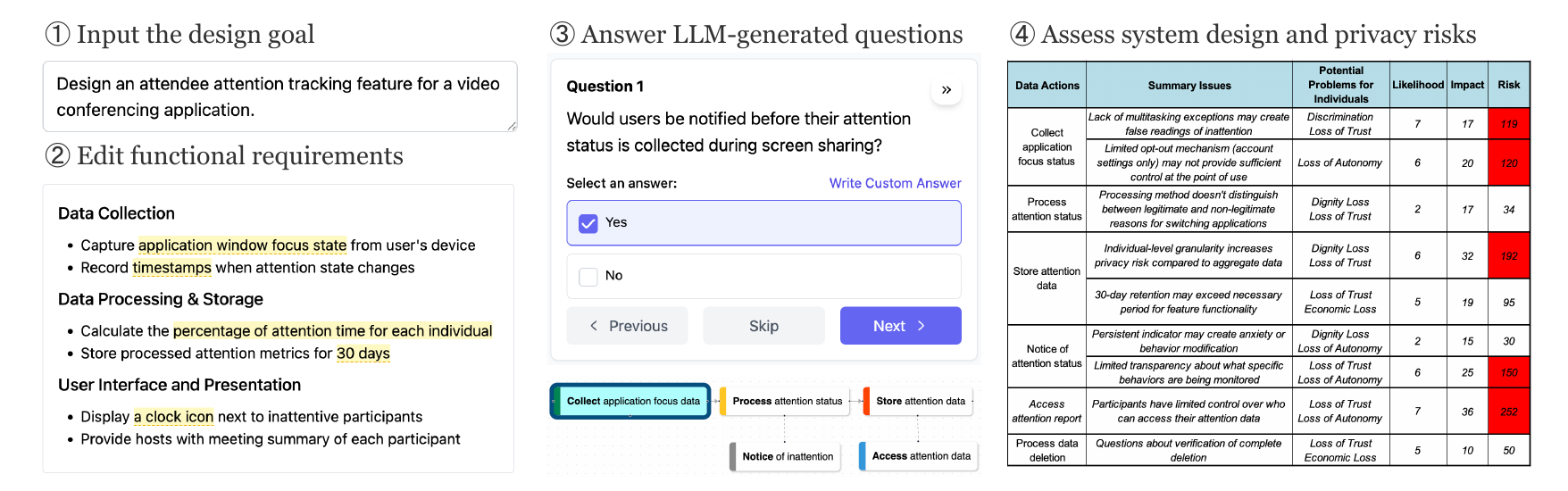}}
    \caption{\name helps developers articulate key privacy design decisions by answering LLM-generated multiple-choice questions. After (1) developers provide a high-level description of their system, (2) \name expands it into detailed, editable functional requirements with adaptable design choices highlighted.  (3) \name guides developers through key design decisions with contextualized, specific questions. As developers make choices, \name organizes and documents these decisions within a structured privacy representation. (4) \name maps these design decisions to NIST PRAM worksheets~\cite{enterprise2020nist} to further identify potential issues and prioritize privacy risks. Note that we only enumerate data actions and relevant design decisions; analysts still need to manually evaluate risk likelihood and impact. 
    }
    \label{fig:overview}
\end{teaserfigure}

\maketitle

\section{Introduction}


The U.S. National Institute of Standards and Technology (NIST) 
released the Privacy Risk Assessment Methodology (PRAM) in 2019 to help organizations analyze, assess, and prioritize privacy risks~\cite{enterprise2020nist}. 
PRAM provides step-by-step guidance on risk assessments through four worksheets (WS): WS1 defines business objectives and organizational privacy governance; WS2 assesses system design; WS3 helps prioritize risks based on likelihood and impact; and WS4 selects appropriate controls to mitigate risks. 
Despite the institutional endorsement, there is little empirical evidence on the effectiveness of NIST's PRAM in practice~\cite{iwaya2024privacy,wijesundara2025sok}. 

In addition to NIST's PRAM, several other frameworks offer complementary strategies to assess privacy risks, such as Privacy Impact Assessment (PIA)~\cite{wright2012privacy} and LINDDUN~\cite{deng2011privacy}. These methods typically follow a similar process: defining system objectives, system mapping, threat enumeration, impact/risk analysis, and mitigation~\cite{khalil2024threat}. However, prior studies found that these frameworks can be cumbersome and overly complex~\cite{wuyts2020linddun, balsa2023technocracy}, often requiring significant expertise to use effectively~\cite{wuyts2014empirical}.  As a result, many organizations, particularly startups without dedicated privacy professionals, perceive privacy risk assessments as onerous burdens rather than practical tools for managing privacy risks~\cite{nistOnline,waldman2021industry}\revise{.} 

In this paper, we investigate methods to lower the barrier for novice developers to apply privacy risk assessments. 
We first performed an observational study to explore the challenges they face when applying PRAM. We recruited 12 participants, including students and junior software engineers, and asked them to conduct privacy risk assessments for four task scenarios using PRAM worksheets. 
We began with a brief training session that walked participants through the four worksheets and demonstrated their use with an example from PRAM materials. Participants then selected one scenario they were familiar with, and followed the worksheet instructions to complete each one.
We found that participants struggled most with articulating privacy design decisions within the system, which revealed three key challenges: (1) participants struggled to distribute their attention across numerous privacy design decisions; (2) PRAM's vague terminology and open-ended structure created confusion about what constitutes appropriate responses; and (3) participants often missed important design decisions without explicit prompting. 

We then designed \name (Figure~\ref{fig:overview}), an interactive tool that helps developers articulate key privacy-related design decisions by answering multiple-choice questions generated by large language models (LLMs). The tool specifically targets PRAM’s WS2 (Assess System Design), as novice developers struggled most with this stage in our studies. \name transforms this open-ended task into a series of multiple-choice questions with concrete answer options.
By decomposing system design into discrete design choices, \name allows developers to focus on one decision at a time. The system leverages LLMs to transform abstract privacy concepts into specific, contextual questions with concrete options. By dynamically generating questions based on previous responses, \name guides developers' attention toward key design decisions that they might otherwise overlook.





While prior studies have explored using question answering to communicate privacy concepts or explain system behaviors~\cite{harkous2016pribots,ravichander2019question,feng2024understanding}, these efforts primarily focus on explaining existing content such as privacy policies. In contrast, the unique design challenge for \name is to guide developers through a vast, underexplored privacy design space. 
\name has three key innovations. 



First, we design a novel universal representation for privacy design (Figure~\ref{fig:verb}) that captures key privacy design decisions to help developers concentrate upon the essentials without distraction from irrelevancies.
Unlike existing models (e.g.,  Data Flow Diagrams, Component
Diagrams~\cite{alhirabi2023parrot}) that focus solely on data flows, such as who is
collecting the data, how it is collected, when, and why, our representation also captures interaction designs with different stakeholders. We organize privacy-related information about a data practice into three levels: (1) a high-level data flow modeling information movement through the system, (2) nodes that represent specific data actions (e.g., collect, process) and stakeholder interactions (e.g., obtaining consent, requesting data deletion), and (3) node properties that specify individual design choices (e.g., opt-in/out).

Second, we introduce a data-driven approach to construct a domain-aware privacy design space (Figure~\ref{fig:taxonomy}). In contrast to prior work that creates taxonomies through manual labeling~\cite{wilson2016creation}, we mine domain-specific taxonomies of privacy design decisions by analyzing privacy-relevant documents, such as public privacy news reports.
Building on our privacy representation, we map each decision to a key-value pair, organized into three categories: (1) universal keys with universal values (e.g., \textit{consent\_mode: opt-in/opt-out}), (2) universal keys with domain-specific values (e.g., \textit{data\_type: medical records}), and (3) domain-specific keys with domain-specific values (e.g., \textit{voice\_masking: enabled}). 
We first reviewed the literature on privacy design~\cite{rubinstein2013privacy, gurses2011engineering, feng2021design, schaub2015design} and manually created an initial design space based on this structure. To enrich this design space, we text-mined 10K privacy-related news using LLMs, annotating domain labels and extracting relevant design choices. We then analyzed their co-occurrence statistics to uncover associations of privacy design decisions across different contexts.




Third, we devise a question generation workflow that dynamically prioritizes key privacy design decisions. \name maintains an underlying representation of the current design and formulates two types of questions: exploratory questions that add new nodes to the data flow, and exploitative questions that leverage the design space to elicit specific design decisions for existing nodes. Since presenting all potential decisions at once would overwhelm developers, \name strategically determines which decisions to present first. In each round, \name selects the question type to balance exploration and exploitation, retrieves relevant design decisions, and prioritizes those strongly correlated with prior choices. 

We conducted two experiments to validate the effectiveness of our design. We first evaluated the coverage of key design decisions using our system. To establish ground truth, we organized brainstorming sessions with privacy experts to examine 30 real-world data practices. We provided initial descriptions of each data practice and answered questions based on their actual implementations.
Our results show that \name identified 94\% of key design decisions recognized by privacy experts, and its options covered 77\% of design choices made in practice. 
We then performed a user study with 24 participants and asked them to apply the NIST PRAM framework to three real-world application scenarios with or without our tool. We found that developers using \name were able to articulate 47\% more key privacy decisions with 73\% less time. 

In this paper, we make the following contributions:
\begin{itemize}[leftmargin=*, noitemsep, topsep=0pt]
    \item An empirical study identifying three key challenges of applying PRAM for novice developers.
    \item \name, a novel tool 
 that uses LLM-generated multiple-choice questions to help developers articulate key privacy design decisions.
    \item A universal representation for privacy design that abstracts privacy-related design decisions into data flows and stakeholder interactions.
    \item A data-driven approach to construct a domain-aware privacy design space by mining privacy-related news.
\end{itemize}

\section{Related Work}
The main objective of this paper is to lower barriers for novice developers to conduct privacy assessments by (1) transforming privacy design into a structured task and (2) guiding developers through the process step-by-step using a series of multiple-choice questions.
We organize related work into three categories: structuring privacy design, developer support for privacy, and privacy question answering.

\subsection{Structuring Privacy Design}
A central challenge in supporting developers with privacy engineering is that privacy design decisions are often diffuse, implicit, and difficult to articulate. We conceptualize structuring privacy design as systematically identifying and organizing privacy-related design decisions. Prior work has approached this problem by developing privacy representations and mapping out the privacy design space.




\sssec{Privacy Representation}. A fundamental challenge in engineering privacy into systems is to articulate privacy~\cite{gurses2011engineering,bokaei2020disambiguating,hosseini2021analyzing,hosseini2021ambiguity}. 
Solove proposed a taxonomy focused on privacy violations~\cite{solove2005taxonomy}. Contextual Integrity~\cite{nissenbaum2004privacy} defines privacy as appropriate information flows. 
However, these representations are too high-level for practitioners to think about concrete system design decisions.

Recent studies found that practitioners often combine general-purpose diagrams to articulate a privacy design~\cite{li2024redesigning, brooks2017introduction, wen2025teaching, tiwari2012merging}, such as Data Flow Diagram (DFD)~\cite{li2009data} and Unified Modeling Language (UML)~\cite{uml2017usecase}. 
Privacy nutrition labels offer a standardized format for disclosing what data is collected and how it is used~\cite{kelley2009nutrition, kelley2010standardizing}. These representations help describe data practices, but they do not explicitly surface the underlying design decisions.

We hypothesize that not all design decisions are equally important; a common subset (e.g., retention periods, consent options) accounts for a substantial share of decisions developers face in practice. Our goal is to design an alternative privacy representation that makes these key decisions explicit and decomposes system design into discrete decisions.

\sssec{Privacy Design Space}. The complexity of privacy design has driven researchers to develop structured approaches for mapping out the privacy design space, particularly for notice and choice mechanisms~\cite{feng2021design, schaub2015design}. For example, Schaub et al. proposed a design space for privacy notices by identifying key dimensions such as timing, modality, and channel~\cite{schaub2015design}. While these efforts provide usable taxonomies and vocabularies for categorizing and communicating different privacy designs, they focus narrowly on notice and user control rather than the broader privacy design landscape.

In parallel, there has been extensive work on the analysis of privacy policies~\cite{zimmeck2014privee, tang2023policygpt, harkous2018polisis, cui2023poligraph, shvartzshnaider2018analyzing}. 
Wilson et al. developed the OPP-115 taxonomy~\cite{wilson2016creation} that categorizes privacy practices by manually annotating a corpus of website privacy policies.
Building on such efforts, natural language processing (NLP) techniques have been widely used to automate the extraction and summarization of salient information from privacy policies~\cite{zimmeck2014privee, liu2014step, ammar2012automatic,bannihatti2020finding,ravichander2019question}. For example, ToS;DR extracts key points from privacy policies and presents them as easy-to-read summaries, improving accessibility and user understanding~\cite{tldrOnline,srinath2024automated}.
However, these approaches offer little support for developers articulating design decisions because privacy policies focus on external-facing high-level statements, rather than implementation-level decisions developers need to consider when designing a system.

We hypothesize that a data-driven approach to constructing the privacy design space can better scale across diverse domains and surface recurring design decisions that matter in practice. In this work, we explore the feasibility of developing a privacy design space by mining design decisions from real-world data practices.

\subsection{Developer Support for Privacy Design}
Prior work has explored various ways to support developers in making privacy design decisions~\cite{li2018coconut, li2024matcha}. Prior studies have identified common challenges developers face in understanding and implementing privacy principles~\cite{tahaei2021privacy, li2021developers, ayalon2017developers, senarath2018developers}. Developers often reported limited privacy awareness and knowledge, particularly in startups and small organizations that lack dedicated privacy experts~\cite{kekulluouglu2023we, prybylo2024evaluating}. Without formal privacy training or institutional support, many developers rely on ad-hoc knowledge sources, which may be inconsistent or incomplete~\cite{prybylo2024evaluating}. Many developers think of privacy in terms of security, missing broader privacy concerns such as data retention and internal misuse~\cite{hadar2018privacy}.

Several tools have been developed to embed privacy practices into software development workflows. For example, Coconut is an Android Studio plugin that helps developers manage privacy through required annotations~\cite{li2018coconut}. PARROT supports privacy-aware IoT development with interactive guidance~\cite{alhirabi2023parrot}. However, existing tools focus on supporting experienced developers with prior privacy knowledge. Little attention has been given to helping novice developers conduct privacy risk assessments. Our work aims to bridge this gap by decomposing the privacy assessment task into an interactive, step-by-step process that guides novices through concrete design decisions.


\subsection{Privacy Question Answering}
Prior research has explored the use of privacy question answering as an effective approach to improve the usability of privacy policies~\cite{feng2024understanding,ravichander2019question,ravichander2021breaking,harkous2016pribots}. For example, PriBots developed conversational assistants that answer users' questions about privacy practices to enhance readability and user comprehension~\cite{harkous2016pribots}.
Recent work in explainable AI explored using LLMs to build dialogue-based privacy tools~\cite{chen2025clear,freiberger2025you,sun2025empowering}. For example, CLEAR used contextual, LLM-powered assistants to analyze privacy policies in real-time, highlight possible risks, and generate explanations as users interact with LLM applications~\cite{chen2025clear}.
Sun et al. built an open-ended QA agent to help users understand privacy policies~\cite{sun2025empowering}. 
In contrast, \name is unique in (1) helping developers rather than users; (2) generating closed-ended questions to elicit design decisions rather than answering open-ended questions by interpreting existing policies.

Privacy Nutrition Labels require developers to answer questions about their app’s data practices~\cite{kelley2009nutrition, kelley2010standardizing}.
For example, Apple Privacy Nutrition Labels ask developers to specify which data types their app collects, how this data is used, and whether it links to user identity or is used for tracking purposes~\cite{kollnig2022goodbye, zhang2022usable}.  However, prior studies have found that filling out these forms is time-consuming and requires significant expertise ~\cite{li2022understanding, flaherty2000privacy}. The labels often use vague terms to accommodate broad use cases, which may confuse developers~\cite{li2022understanding}. 
In contrast, our work aims to help novice developers articulate privacy design decisions using multiple-choice questions similar to the Akinator game~\cite{sasson2023mirror}. By leveraging LLMs, we generate specific, contextualized questions with concrete options to reduce ambiguity and potential confusion.

 

\section{Background}
In this section, we review existing techniques for privacy risk assessments and motivate our choice of PRAM.

\sssec{Privacy Impact Assessment (PIA)}~\cite{wright2012privacy} is a process used to evaluate the potential effects of information systems on individual privacy. PIAs are increasingly adopted by government agencies and organizations to encourage early integration of privacy considerations into the system development lifecycle~\cite{wright2012state,iwaya2024privacy}.
However, prior research found that existing PIA processes typically lack clear guidelines or methodologies to sufficiently support privacy risk assessments (PRA)--the technical evaluation of privacy risks~\cite{de2016priam, alshammari2018towards}. Consequently, PIAs often rely on ad-hoc, checklist-like approaches that serve more as compliance rituals than actually addressing privacy risks~\cite{waldman2021industry, clarke2009privacy,balsa2023technocracy}.

Several efforts have attempted to address these gaps~\cite{de2016priam, alshammari2018towards, deng2011privacy,oetzel2014systematic}. For example, PRIAM introduces a risk model that formalizes key risk factors: privacy harms, feared events, privacy weaknesses, and risk sources~\cite{de2016priam}. LINDDUN develops a taxonomy of privacy threats for identifying and mitigating privacy risks~\cite{deng2011privacy}.
However, these frameworks are often overly abstract or complex, requiring significant privacy expertise to implement effectively~\cite{wuyts2020linddun, sion2025robust, wijesundara2025sok}.











\sssec{Privacy Risk Assessment Methodology (PRAM)}~\cite{enterprise2020nist} stands as a notable effort to provide systematic guidance for conducting PRA. PRAM instantiates the risk model from NISTIR 8062~\cite{brooks2017introduction}, which defines risks as the product of likelihood and impact of adverse privacy effects for individuals. PRAM guides practitioners through PRA using four worksheets (WS), each containing a set of tasks:


\begin{itemize}[leftmargin=*]
    \item \textit{WS1: Framing Business Objectives and Organizational Privacy Governance}. WS1 requires analysts to elicit system requirements, including functional requirements (e.g., business goals) and privacy-related non-functional requirements (e.g., legal obligations).
    \item \textit{WS2: Assessing System Design}. WS2 focuses on privacy threat modeling. \textit{Task 2 (Supporting Data Map)} instructs analysts to create a system model or data flow diagram.
    \textit{Task 3 (Contextual Factors)} encourages analysts to consider contextual factors such as the nature of the organizations and privacy expectations of these organizations.
    \textit{Task 4 (Data Action Analysis)} requires the analyst to fill out a table by enumerating data actions in the system, the data being processed, relevant contextual factors, and a summary of issues.
    \item \textit{WS3: Prioritizing Risk}. WS3 comprises four tasks: (1) \textit{Assess Likelihood}: estimating the probability that a data action will become problematic for individuals, (2) \textit{Assess Impact}: estimating the effects of potential problems for individuals using the organizational impact factors, (3) \textit{Calculate Risk}: multiplying likelihood and impact to produce a risk score, and (4) \textit{Prioritize Risks}: providing suggestions to help the organization prioritize risks.
    \item \textit{WS4: Selecting Controls}. WS4 supports the selection of controls to mitigate privacy risks identified in WS3.
\end{itemize}


With the proliferation of digital services (e.g., mini-programs and microservices) and the rise of AI programming tools, software is increasingly being developed by individuals or small teams who lack privacy expertise~\cite{prybylo2024evaluating, brutschy2014static, dohmke2023seachangesoftwaredevelopment, kekulluouglu2023we}. However, existing privacy risk assessment frameworks are mostly not designed for novices, and can be challenging for them to apply effectively.
Based on these gaps, we propose the following research questions:

\textit{RQ1: How can privacy design be turned into a structured task?}

\textit{RQ2: How can the structured privacy-assessment task be unfolded for novice developers?}

\section{Understanding Challenges of Using PRAM for Novice Developers}
\label{sec:nist_study}
We conducted a study with 12 participants to explore the challenges of applying privacy risk assessments for novice developers. We used PRAM as a need-finding tool to uncover developers’ underlying needs and pain points. We selected PRAM over other frameworks since (1) NIST offers well-structured worksheets and instructional materials; (2) PRAM is more lightweight than LINDDUN~\cite{deng2011privacy} and LINDDUN Go~\cite{wuyts2020linddun}, which better suits novices.

\begin{table*}[htbp] 
\centering\small
\caption{Task scenarios used in our study. We selected four task scenarios based on participant familiarity, scenario complexity, and domain diversity. For each scenario, we provided participants with a high-level description of the system.
} 
\Description{A table listing four scenarios used to study how developers conduct privacy risk assessments. The columns are ID, Scenario (e.g., "Zoom Attendee Attention Tracking"), Description, and Participants (showing which study participants were assigned to each scenario).
}
\begin{tabular}{c p{33mm} p{110mm} p{16mm}}
\toprule 
\textbf{\#ID} & \textbf{Scenario} & \textbf{Description}  & \textbf{Participants}\\ 
\midrule
\#1 & Zoom Attendee Attention Tracking~\cite{li2024redesigning, Amatulli_2020} & A feature that monitored attendees' attention during video conferences by tracking whether Zoom was the active application on a participant's computer. It showed a clock icon next to inattentive participants and generated a post-meeting report with attention percentage scores. &  P1, P2, P3, P8\\ 
\midrule 
\#2 & Alexa's Smart Home Voice Assistant~\cite{Matt_Day_2019} & A cloud-based voice service for smart home devices that allows users to control their home environment through voice commands. It captures audio when activated by a wake word, processes requests in the cloud, and executes actions through connected devices. &  P4, P7, P11\\ 
\midrule 
\#3 & Target Retail Recommendation System~\cite{Hill_2024} & An analytics system that identified pregnant customers based on purchase patterns of approximately 25 indicator products, allowing Target to predict pregnancy and estimate due dates. This enabled targeted marketing of pregnancy and baby-related products. &  P5, P9, P12\\ 
\midrule 
\#4 & 23andMe Direct-to-consumer Genetic Testing~\cite{Mineo_2025} & A service that collects and analyzes genetic information from customer saliva samples for ancestry, health traits, and disease risks. The service stores comprehensive genetic profiles that can reveal sensitive information about individuals and their biological relatives.  & P6, P10\\ 
\bottomrule 
\end{tabular} 
\label{tab:nist_study}
\end{table*}



\subsection{Method}
\label{sec:study_method}
\subsubsection{Recruitment} Since our use case envisions startups hiring new graduates without dedicated privacy professionals, we primarily recruited students and junior professionals. We posted the recruitment advertisements through social media, including our institution’s subreddit/Slack/Discord channels, and student mail lists, and sent targeted invitations to junior software developers.

Participants who expressed interest in the study were asked to fill out screening forms, which we used to determine their eligibility for the study. Selection criteria were based on age (over 18 years) and software development experience (e.g., coursework, internships, or professional work). During screening, we excluded participants who had formal industry or research experience in privacy engineering, as the studies targeted novices lacking privacy expertise.

\subsubsection{Participants} 
We recruited 12 participants (five identified as female, seven as male, aged 22 to 24) through social media and mailing
lists. Each participant was compensated for their time with a \$20 gift card. The sample included 2 graduate students, 9 undergraduates, and 1 software engineer. To assess participants' privacy knowledge, we asked them to rate their familiarity with privacy principles or ``Privacy by Design'' practices on a scale from 1 (not familiar) to 5 (very familiar). Reported knowledge was low (M = 1.58, SD = 1.11), with 10 out of 12 scoring 2 or below.




\subsubsection{Study Procedure} We assumed participants to be the developers responsible for designing a specific app or service and asked them to use PRAM worksheets~\cite{enterprise2020nist} to assess the privacy risks of one data practice. 
We first gathered a broad range of privacy-related scenarios, then selected four to cover diverse privacy-related design dimensions (e.g., interface, cloud, AI, and physical) across different domains (e.g., video conferencing, IoT, advertising, and health). We also chose scenarios with moderate complexity, so they do not impose a huge understanding overhead on participants. We asked participants to select one of four data practices (Table~\ref{tab:nist_study}).  For each scenario, we provided only a high-level description of its functional requirements to reflect real-world conditions.
We imposed a 2-hour time limit for the study. We captured participants' responses through a hybrid format: hand-written format for WS1 and WS2 to facilitate brainstorming and drawing tasks, and using the NIST open source digital version (Excel) of WS3 to calculate risk. The choice among four data practices was selected based on participants' interests and familiarity with the cases, allowing them to engage with scenarios they found relevant to their experience. After the study, three authors
reviewed the participants’ assessments and asked additional
questions to clarify details in their assessments.
We started with broad questions to ask about participants’ overall experiences and challenges with PRAM, followed by more targeted questions about their decision-making on specific worksheet tasks.
We conducted all study sessions in person to allow for detailed observation and immediate feedback.


\subsubsection{Data Analysis} We conducted a thematic analysis~\cite{wicks2017coding} of post-task interview transcripts and notes we took during the study. Two researchers independently coded 8 transcripts, then discussed their codes and developed a codebook that the two researchers agreed on. Using this codebook, they divided the remaining transcripts, with each coding two transcripts. Finally, the two researchers discussed and resolved coding conflicts. As our focus was on identifying emerging themes, we did not calculate inter-rater reliability (IRR) to measure theoretical agreement.

\subsubsection{Ethical Considerations} The study received approval from the Institutional Review Board (IRB) of our institution. We obtained participants’
consent electronically via Google Forms at the beginning of each study, and reserved their rights to withdraw at any time in the study. We informed participants that the information collected during the interview will be anonymized and only the results obtained from analyzing the transcribed interviews will be disclosed. To ensure participants’ privacy, we used Zoom (approved by our institute) for transcription. After each interview, the research team checked the transcription using the recording, then immediately and permanently deleted the recording.

\subsection{Findings}
We identified four main challenges of conducting privacy risk assessments for novice developers based on participants' behaviors and feedback. The first three challenges relate to how they articulate system design, while the fourth involves estimating risk scores for privacy risks. We provide the codebook in Appendix~\ref{sec:codebook}.




\subsubsection{Attention Allocation} Participants struggled to manage their mental resources to focus on important design considerations. We identified two main causes of this challenge.

\sssec{Information Overload}.
The complexity of PRAM led to cognitive overload for all participants, who had to juggle privacy concepts, technical details, and organizational factors simultaneously. As sessions progressed, this overload led to noticeable fatigue. Several participants (P1, P2, P4, P5, P7) described struggling to ``keep all of this in [their] head at once'' despite repeated exposure.

This cognitive strain was reflected in their task performance. Several participants offered detailed responses in WS1, but failed to articulate the same level of detail in WS2. In post-task interviews, participants attributed this drop-off to mental overload, with P4 noting that it was ``\textit{too hard to recall and include all the details...in one graph}.''

\sssec{Ineffective Prioritization} (P1, P3, P4, P5, P7, P8, P11). Many participants struggled to prioritize privacy considerations effectively, often focusing on technically familiar aspects rather than the most critical privacy risks. Without clear guidance, they misallocated attention, spending too much time on early, familiar tasks and giving insufficient attention to later, equally important ones.

This was particularly evident in WS2 (\textit{Assessing System Design}). Several participants concentrated heavily on the first data action (e.g., collection), but included only brief or incomplete responses for later actions like retention, transformation, or disclosure. P5 reflected that "\textit{I spent too much focus and time on the first data action}."
In task \#1, P8 spent 10 minutes on collection but rushed through the rest, missing key issues like third-party disclosure. P8 explained that ``\textit{By the time I reached disclosure, my focus was fragmented}.''

Participants tended to focus on privacy design decisions that aligned with their technical background, often giving less attention to unfamiliar but equally important aspects. For example, P4, a software engineer, provided detailed input on storage but overlooked the frequency of wake word detection, an important decision related to household surveillance. P4 explained that ``\textit{I wasn't confident about the low-level detail, so I chose not to extend that option}.'' In other cases, participants defaulted to whatever came to mind, with P11 noting that “\textit{There are too many variables...I just couldn't think that more}.”

\subsubsection{Inaccessible Privacy Knowledge}
 Participants often lacked the knowledge of what privacy aspects to consider or failed to recall relevant concepts when needed. 
 
    
    \sssec{Knowledge Blindspot} (P3, P8, P11). Several participants missed important design decisions because they did not know what to consider. In the Zoom attention tracking scenario, P8 initially overlooked issues such as data deletion and access control. However, during the post-study interview, when these considerations were discussed, he immediately recognized their importance and suggested improvements like allowing users to control deletion timing and limiting access to attention reports. P11 noted that ``\textit{It is not intuitive to do a systematic analysis of this task if I don't have too much privacy knowledge}.''

    \sssec{Knowledge Recall Failure} (P1, P2, P4, P7, P8).
    In several cases, participants had relevant privacy knowledge but didn’t apply it when needed, often due to memory lapses or misdirected attention.
Many Participants  (P1, P2, P4, P5, and P7) struggled to carry earlier insights into later tasks. P5 noted that they had ``\textit{focused too much on the first part},'' leaving little mental capacity for later stages. In the Zoom scenario, some participants overlooked user consent mechanisms, noting that they had considered them earlier but forgot to include them during the assessment because ``[PRAM] \textit{never prompted [them] to consider it}''.
P8 shared a similar experience, explaining that he forgot to include a notification ``\textit{not because he didn’t know it, but because it slipped my mind}'' during the task. He only recalled the information when the researcher brought it up in the post-study discussion.

\subsubsection{Insufficient Guidance} Participants found PRAM’s instructions vague, with unclear expectations and ambiguous terms that left them uncertain about the level of detail and what to include.

    
    \sssec{Unclear Expectations for Detail and Scope} (P2, P4, P5, P7).
    Participants were often unsure how much technical depth was expected in their responses. The framework lacked examples to guide the level of specificity, leading to confusion for participants. P4 directly expressed this uncertainty: ``\textit{Am I supposed to break down this system to the level of specific algorithms and data structures, or keep it high-level?}''  This lack of clarity led participants to over-focus on their familiar areas, while overlooking critical issues such as user consent.
    
    Beyond ambiguity in granularity required, participants also felt uncertain about what to include and how comprehensive their assessments should be. 
    Many participants felt ``\textit{unsure what’s expected},'' and questioned whether they had done ``\textit{enough}'',   which led to hesitation and second-guessing. P7 mentioned that ``\textit{I just wish there are more specific instructions on what to fill out}.''

    \sssec{Vague Terms} (P3, P4, P5, P7, P9, P12).
    Participants felt that the terminology in PRAM WS2 was unclear. Terms used in PRAM (e.g., ``generation/transformation'' and ``disclosure/transfer'') were perceived as ``\textit{overly technical}'' or ``\textit{ambiguous}''. As P12 stated, ``\textit{The format is unclear and needs clearer definitions, for example, the difference between `transformation' and `transfer'}.''

This confusion extended to the definitions and structure of the worksheets. Several participants noted inconsistencies in how concepts were presented.  For example, the term ``contextual factors'' is initially defined as ``circumstances surrounding the data by the system or individuals that influence whether a data action may be problematic'', with examples like data sensitivity, data collection frequency. However, in the Contextual Factors tab, participants encountered a different framing--subjective factors from organizational, system, and user perspectives, such as user perceptions of data sensitivity. This mismatch left the participants unsure how to respond. As P12 noted, ``\textit{I wasn’t sure what to fill out because the examples didn’t match.}''

\subsubsection{Uncertainty in Risk Evaluation}
Participants encountered two main challenges when applying PRAM’s risk assessment approach in WS3: difficulty assigning risk scores and uncertainty about how to interpret them.

    
    \sssec{Inconsistent Risk Scoring} (P3, P8). PRAM required analysts to independently evaluate each potential privacy problem across multiple organizational impact dimensions (e.g., compliance, business, reputation, culture) and assign 1-10 ratings to each factor. 
    We observed significant variation in how participants rated similar privacy issues due to unclear guidance on how to weigh impact categories. For the same scenario, P3 gave a low score for surveillance-related autonomy loss (7), while P8 rated it much higher (26). A similar gap appeared in their dignity-related assessments (6 vs. 17). As P3 noted, ``\textit{I wasn't sure how to weigh different kinds of organizational impacts against each other. Does a regulatory fine matter more than losing customers' trust? The framework doesn't tell you.}''

    \sssec{Unclear Meaning of Risk Scores} (P5, P11, P12).
    Even after assigning scores, participants struggled to interpret their meaning. PRAM does not define thresholds for high, medium, or low privacy risks, leaving participants to set arbitrary cutoffs. In task \#3, participants calculated significantly different risk scores for different data actions (ranging from 152 to 739), but struggled to determine which risks required action due to the lack of guidance on how to interpret these scores (P5, P11, P12). P5 expressed confusion about the resulting score, ``\textit{Without some kind of baseline, these numbers don't really tell me what that means to my scenario}''. Without clear thresholds, participants had difficulty prioritizing risks confidently or consistently.

\section{\name Design Overview}

This section presents the high-level design of \name, including design goals, architecture and system scope.

\sssec{Design Goals}. Motivated by the challenges identified in our observational study, we derive the following design decisions to address the most pressing user needs (i.e., articulating system design).

\begin{itemize}[leftmargin=*]
    \item \textit{Information overload}: Drawing on distributed cognition theory~\cite{norman2014things, hollan2000distributed}, we design a structured representation as an external artifact~\cite{norman2014things} to support thinking and offload cognitive load. A good representation captures the key privacy design decisions while deliberately leaving out the rest, allowing developers to only focus on what matters.
    \item \textit{Ineffective prioritization}: Novices often stick to the first thing they notice, so we explicitly prompt them to consider different privacy aspects and use a mix of explorative and exploitative questions to balance breadth and depth. 
    \item \textit{Knowledge blindspots}: We mine privacy-related news to build a knowledge base of key privacy design decisions. 
    \item \textit{Knowledge retrieval failure}: Our representation enables navigation across detailed privacy design decisions while maintaining a global view. We also use proactive prompts to help recall relevant privacy knowledge.
    \item \textit{Insufficient guidance}: We transform the open-ended risk assessment process into a series of multiple-choice question answering to produce a more structured and guided workflow.
    \item \textit{Vague terms}: We leverage LLMs to ground questions in the developer’s context and provide concrete answer options.
\end{itemize}

\sssec{System Architecture}. A key design of \name is to use multiple-choice question answering to articulate privacy-relevant design decisions. Instead of filling out open-ended forms, developers respond to a sequence of contextual, specific questions that help identify and document critical privacy choices.  
In doing so, \name guides developers in thinking through the privacy design space by breaking it down into manageable, focused steps tailored to their system context.

\name comprises three components: (1) a \textbf{universal privacy design representation} that enables systematic organization and documentation of privacy-relevant design decisions (Section~\ref{sec:representation}), (2) a \textbf{domain-aware privacy design space} that uncovers privacy design choices across domain-specific contexts (Section~\ref{sec:design_space}), and (3) a \textbf{dynamic question generation workflow}  that prioritizes key design decisions based on the developer’s previous answers and system context (Section~\ref{sec:question_generation}). 



These components work together to generate questions that progressively explore the privacy design space. \name draws from the design space to create contextually appropriate questions, prioritizing key design decisions most relevant to the developer’s system. As developers respond to the questions, their answers incrementally enrich the underlying privacy representation.

\name employs LLMs to extract information from news and user inputs, generate choices from embedded knowledge, and translate structured representations into natural language. These capabilities—extraction, recall, and generation—are among the most mature across LLMs~\cite{chang2024survey}. We intentionally minimize the use of LLM reasoning. \revise{We further discuss how \name mitigates the risks of relying on LLM-generated content in Section~\ref{sec:discussion}.}


\sssec{System Scope}. \name is a human-in-the-loop tool that helps developers articulate the key design decisions.
Specifically, \name assists with PRAM's \textit{WS2: Assess System Design},
as our previous studies (Section~\ref{sec:nist_study}) suggest that novice developers struggled most with this system design articulation process.
Additionally, developers cannot effectively engage with later PRA stages (e.g., prioritizing risks) without clearly identifying and describing the specific design decisions first. \name provides hints to help novice developers consider decisions they may have overlooked, but it does not guarantee full coverage.





While our studies also reveal broader challenges in later risk assessment phases, such as subjectivity and a lack of standards in risk evaluation, we consider these issues beyond the scope of our focus. Several approaches have been proposed to address these challenges, including quantitative risk models~\cite{luna2012privacy, cronk2021quantitative} and formal methods for risk calculation~\cite{vukovic2014user}, which are complementary to our contribution. We discuss future directions for extending support for risk assessment in Section~\ref{sec:discussion}.

\section{Privacy Representation Design}
\label{sec:representation}

We drew on the theory of distributed cognition~\cite{norman2014things,hollan2000distributed} and iteratively designed a representation for low-level privacy-related design decisions to manage information overload and support effective prioritization. In distributed cognition theory, representations serve as \textit{external artifacts} that offload cognitive work into the environment, so people don’t need to hold all the information in mind simultaneously. Norman similarly defines representations as the way information is structured or presented so that it can be perceived, understood, and used effectively by people, emphasizing that representations are inherently \textbf{selective} in capturing essential elements while deliberately omitting the rest~\cite{norman2014things}.

\begin{figure*}[htbp]
    \centering
    \includegraphics[width=\linewidth]{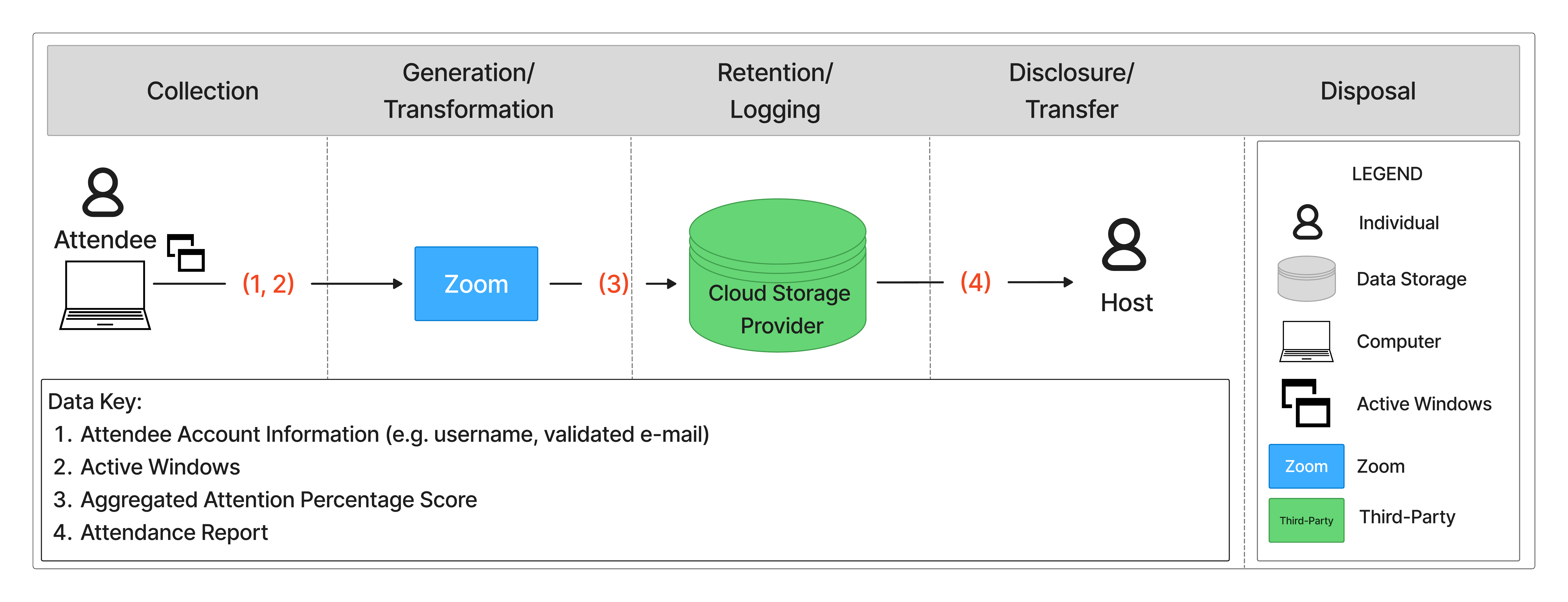}
    \vspace{-15pt}
    \caption{Data Flow Diagram defines data actions that occur across the data lifecycle, indicates the key entities involved in the system, and connects the data flows between them. However, the level of detail is insufficient to capture privacy nuances, and it omits interaction with different stakeholders, and may lead to deferred responsibility for privacy. }
    \label{fig:dfd}
    \Description{A flowchart illustrating the data lifecycle for a Zoom attention tracking feature. An "Attendee" icon is connected to a "Zoom" icon, which then connects to a "Cloud Storage Provider", and finally to a "Host" icon. Numbered data keys indicate what information is transferred at each step: (1) Attendee Account Information, (2) Active Windows, (3) Aggregated Attention Percentage Score, and (4) Attendance Report. A legend explains the icons for Individual, Data Storage, Computer, Active Windows, Zoom, and Third-Party.}
\end{figure*}

Guided by these principles, we conceptualize privacy design around two fundamental components of data systems: data flows and stakeholders. We organize privacy decisions into building blocks that articulate how data moves through a system and how various stakeholders interact with these flows. This representation helps novices focus their attention on key privacy design choices instead of being overwhelmed by the full design space.

We postulate that an ideal representation for privacy design should satisfy the following properties.
\begin{itemize}[leftmargin=*]
    \item \textbf{Concise}: The representation should illustrate privacy-related design decisions concisely.
    \item \textbf{Expressive}: The representation should be expressive to encompass all critical aspects of privacy. 
    \item \textbf{Simple}: The representation should be simple to minimize the learning curve and cognitive load.
    \item \textbf{Objective}: The representation should only present how stakeholders interact with the privacy design but not make any judgment. Evaluation should be independent of the representation.
\end{itemize}

\subsection{Design Iterations}
\label{sec:design_iterations}
We used a bottom-up approach to guide the design of privacy representation. 
We collected 40 real-world data practices from news reports, privacy policies, and literature on privacy design~\cite{rubinstein2013privacy, gurses2011engineering,wilson2016creation,jin2021lean,feng2021design,schaub2015design}, and examined the privacy-relevant design decisions that emerge from these documents. We used similar selection criteria as in Section~\ref{sec:study_method}, but complexity was not considered.
We then designed a representation to accommodate these data practices, iterated on the representations as we expanded the supported use cases, and collected early feedback from software developers through the authors’ personal network.

\sssec{Data Flow Diagram (DFD)}. We initially applied the data flow diagram from PRAM (\textit{WS2: Supporting Data Map}), which visualizes how data moves through the system (Figure~\ref{fig:dfd}). PRAM defines a set of key information system operations, referred to as data actions, that occur across the data lifecycle: \textit{collection}, \textit{generation/transformation}, \textit{retention/logging}, \textit{disclosure/transfer}, and \textit{disposal}.
The diagram identifies the key entities involved in the system (e.g., user, third-party service) and maps the data flows between them.  Each flow is annotated with the specific types of data being exchanged (e.g., names, addresses).


As we tested this representation on more use cases, we observed a few trade-offs:

\begin{itemize}[itemsep=0pt, leftmargin=*]
\item[\textbf{+}] This representation is intuitive and easy to understand.
\item [\textbf{--}] The level of detail is insufficient to capture privacy nuances in data practices (e.g., frequency of data collection).
\item [\textbf{--}] The representation design omits interactions with different stakeholders, such as showing privacy notices or providing user control mechanisms.
\item [\textbf{--}] Many system components fall outside developers' direct control, leading to deferred responsibility for privacy.  For example, while external storage may seem out of reach, choosing third-party services versus self-hosting involves privacy implications that developers should consider.
\end{itemize}

\begin{figure*}[htbp]
    \hspace{-10pt}
    \includegraphics[width=1.02\linewidth]{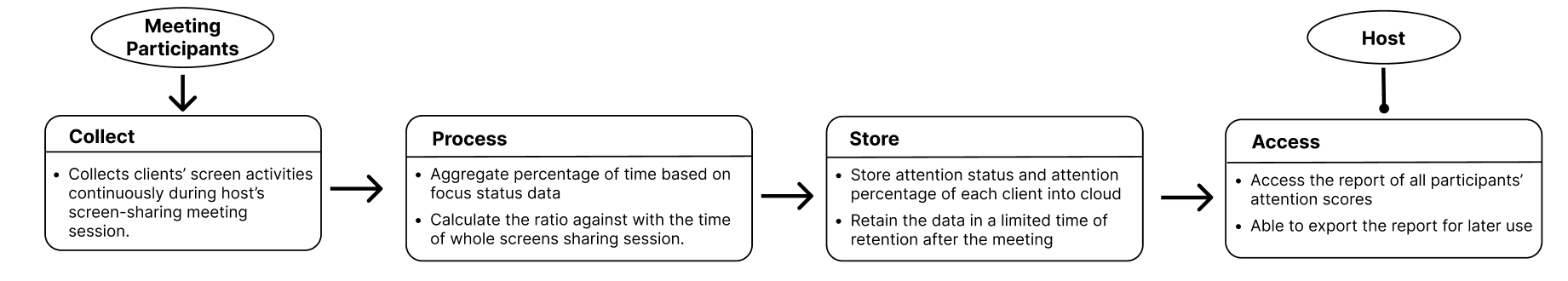}
    \caption{Privacy Storyboard illustrates data actions across the data lifecycle with stakeholder roles. While more aligned with developer workflow, it can overwhelm users with too much information and often conflates multiple design decisions.}
    \label{fig:lpr}
    \Description{A flowchart showing four main stages: Collect, Process, Store, and Access, with arrows indicating the flow from "Meeting Participants" to a "Host". Each stage has a box with a brief description. For example, the "Collect" stage "Collects clients' screen activities continuously during host's screen-sharing meeting session." The storyboard illustrates data actions and stakeholder roles.}
    \vspace{15pt}
\end{figure*}

\begin{figure*}[htbp]
    \hspace{-15pt}
    \begin{subfigure}[b]{0.7\textwidth}
        \centering
        \includegraphics[width=1.1\linewidth]{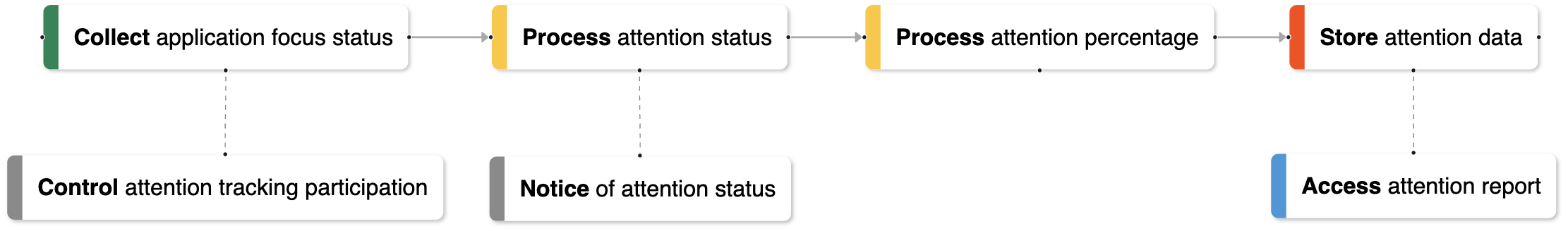}
        \caption{High-level view}
    \end{subfigure}
    \hspace{40pt}
    \begin{subfigure}[b]{0.2\textwidth}
        \includegraphics[width=1.2\linewidth]{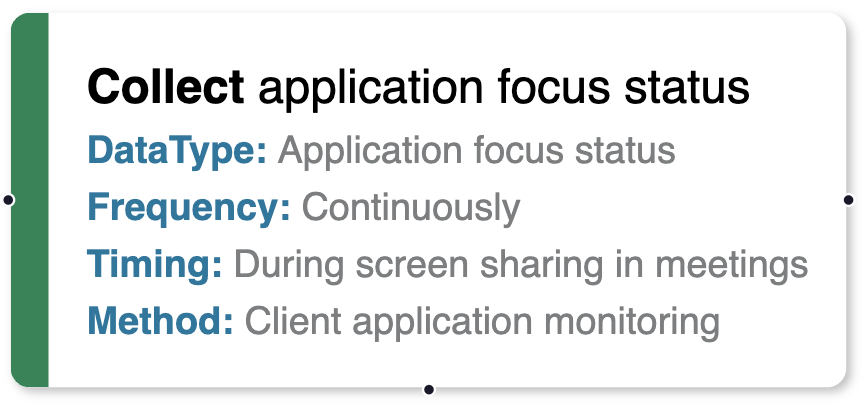}
        \caption{Detailed view}
    \end{subfigure}
    \caption{Our final multi-layer graphical representation adopts a three-layer representation with data flow, stakeholder interaction, and individual design decision. The combination of both a high-level overview and a detailed view supports quick updates and maintaining awareness of the big picture.}
    \label{fig:verb}
    \Description{A diagram showing two views of a data flow. (a) High-level view: A sequence of colored rectangular nodes representing data actions like "Collect application focus status" and stakeholder interactions like "Control attention tracking participation." (b) Detailed view: A pop-up box for the "Collect application focus status" node, showing its properties as key-value pairs, such as DataType: Application focus status and Frequency: Continuously. This demonstrates the tool's ability to zoom in from an overview to specific details.}
\end{figure*}

\sssec{Privacy Storyboard}. We then made three changes to address the limitations of the data flow diagram: (1) shifting to a process-oriented rather than device-focused view to better align with developer workflows, (2) describing implementation of each data action to provide more fine-grained details, and (3) incorporating stakeholder roles to specify who is involved in or can control each part of the data flow. Figure~\ref{fig:lpr} illustrates how participant screen activity is collected, processed, stored, and accessed in \textit{Zoom Attendee Attention Tracking} scenario (Table~\ref{tab:nist_study}), with each step outlining specific operations performed and the stakeholder roles (e.g., attendees giving consent to data collection, host accessing attention reports). 
We observed a few trade-offs in this iteration:
\begin{itemize}[itemsep=0pt, leftmargin=*]
\item[\textbf{+}] The representation design is developer-centric, closely reflecting how developers conceptualize and build systems.
\item[\textbf{--}] The representation presents too much information simultaneously, which may overwhelm readers and make it hard to maintain an overview of the entire privacy design.
\item[\textbf{--}] The representation often conflates multiple design decisions in a single description, making it difficult to extract specific privacy choices for analysis.
\item[\textbf{--}] Merely mentioning stakeholders is too abstract; specific interaction types (e.g., consent or control) are critical for assessing privacy risks but remain inadequately defined.
\end{itemize}




\subsection{Multi-layer Graphical Representation}
As we iterated with more privacy representations, we explored the design space of the data practices. We found that stakeholders only have enumerable ways to interact with a data flow, and abstracted these into a common set of stakeholder interactions, summarized in \revise{Appendix} Table~\ref{tab:definition}.

We then developed a multi-layer graphical
representation for privacy design decisions. Inspired by DENIM~\cite{lin2000denim}, we adopted a three-layer representation similar to website design~\cite{wen2025teaching}: data flow, stakeholder interactions, and individual design decisions (Figure~\ref{fig:verb}). The data flow is organized as a sequence of data actions  (e.g., collect, process, store) connected in a graph structure. Each data action and stakeholder interaction is modeled as an individual node, with stakeholder interactions linked to the relevant data actions. Specific design decisions are captured as node properties, such as frequency and timing of data collection, and collected data types.


The representation offers both a high-level overview and a detailed view. The overview includes only the top two layers (data actions and stakeholder interactions) but omits individual design choices. Users can zoom in on a specific node to see detailed design decisions as key-value pairs.
This approach supports quick updates and helps maintain awareness of the big picture. It allows analysts to outline the overall privacy design before diving into specifics and refer back to the overview to maintain a clear sense of direction throughout their articulation process.

\section{Mining Domain-aware Privacy Design Space}
\label{sec:design_space}

Inspired by Augur~\cite{fast2016augur}, we developed a data-driven approach to construct a domain-aware privacy design space by mining privacy-related news. In this section, we first describe the taxonomic structure of our privacy design space, then present our method for extracting privacy design decisions from privacy news.

\begin{figure*}[htbp]
    \centering
    \includegraphics[width=1.0\linewidth]{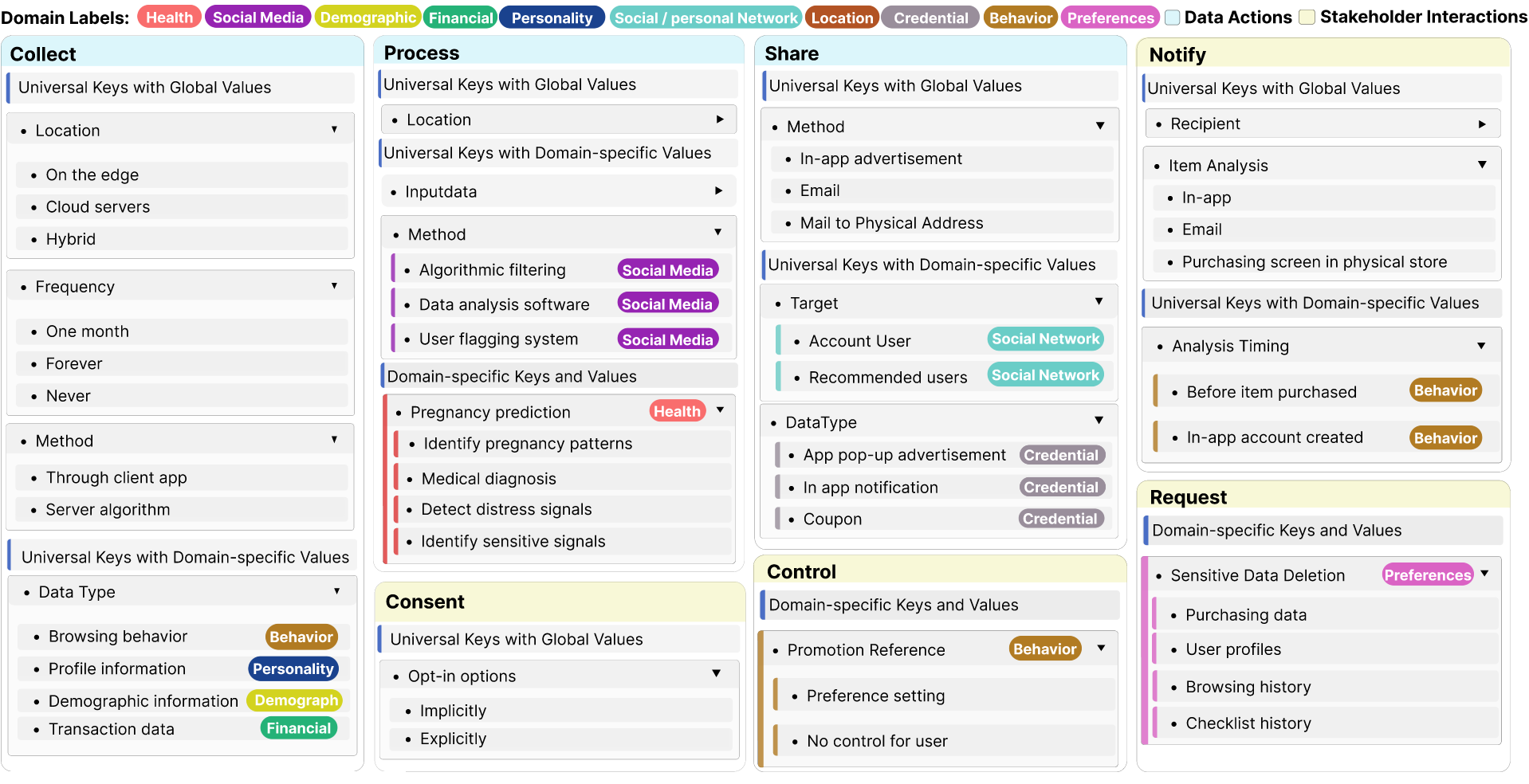}
    \caption{We organize our domain-aware privacy design space by data actions and stakeholder interactions, categorizing privacy design decisions based on the applicability of their keys and values across different domains.
    }
    \label{fig:taxonomy}
    \Description{A large, complex diagram showing a taxonomy of privacy design decisions organized by data actions (Collect, Process, Share, Notify) and stakeholder interactions (Consent, Control, Request). Within each category, decisions are classified as "Universal Keys with Global Values," "Universal Keys with Domain-specific Values," or "Domain-specific Keys and Values," with color-coded domain labels like Health, Social Media, and Behavior.}
\end{figure*}

\subsection{Structuring Privacy Design Space}
\label{sec:design_space_structure}

To understand the structure of the privacy design space, we examined the data practices collected in Section~\ref{sec:design_iterations} using our privacy representation. We used an iterative, open-coding approach~\cite{waldherr2019inductive} to analyze the privacy design decisions. For each design decision, two authors independently annotated its key (the type of decision), the corresponding value (the specific choice made), and the associated data action or stakeholder interaction. We then collaboratively synthesized these openly generated annotations into a coding scheme by merging design decisions of the same type, and re-coded all design decisions using the finalized coding. This process yielded 96 unique privacy design decisions.



From this analysis, we make two key observations. First, we found that different data practices often involve common types of design decisions (e.g., \texttt{data\_type}, \texttt{consent\_mode}, \texttt{storage\_location}).
Second, while most decision types (or keys) are applicable across domains (e.g., health, smart homes), their specific values often depend on the context of data practices. For example, \texttt{data\_type} may involve \texttt{medical\_records} in a healthcare setting, but \texttt{GPS\_coordinates} in a location-tracking app or \texttt{audio\_commands} in a smart speaker. This demonstrates the need for a domain-aware design space that captures context-specific privacy nuances.



Building on our privacy representation, we formulate the structure of the design space by organizing privacy design decisions into three categories (Figure~\ref{fig:taxonomy}): (1) universal keys with universal values that apply across domains (e.g., \textit{consent\_mode: opt-in/opt-out}); (2) universal keys with domain-specific values, where categories remain consistent but options vary by context (e.g., \textit{data\_type: medical records}); and (3) domain-specific keys for decisions unique to particular domains (e.g., \textit{voice\_masking: enabled}).





\subsection{Mining Design Decisions from Privacy News}
\label{sec:mine_design_space}
We first manually built an initial privacy design space of design decisions based on prior analysis results from Section~\ref{sec:design_space_structure} to establish the structure of the design space. 
We then extracted key privacy design decisions from the online corpus of privacy-related news using LLMs to populate and expand this design space. We hypothesize that important privacy design decisions are more likely to be reported in news and incident coverage because mistakes in these decisions tend to have severe consequences for individuals or organizations, triggering public scrutiny and media attention.

\sssec{Dataset Preparation}. We collected 10K news labeled with privacy category from popular technology news sources, including TechCrunch~\cite{TechCrunch}, Guardian~\cite{The_Guardian}, Wired~\cite{Nast}, and The Verge~\cite{The_Verge}. We applied keyword-based filtering using terms like ``privacy'' and ``data protection'' to identify privacy-related news articles.
To enable more precise filtering at scale, we sampled a small set of news articles and used the GPT-4o model to assess their relevance to privacy design. Using these LLM-generated labels, we trained a specialized classifier based on the RoBERTa model~\cite{liu2019roberta} as a cost-effective alternative to LLMs. We then applied this classifier to filter the entire corpus, which yielded 7,058 news articles related to privacy design.


\sssec{Annotating Domain Labels}. For each filtered news article, we segmented the news text into individual data practices to isolate distinct privacy scenarios. We then annotated each data practice with a set of domain labels derived from Privacy Contextual Domains in MITRE's PANOPTIC Taxonomy~\cite{katcherpanoptic}, which consists of 20 labels covering both data types and application domains (e.g., financial, health).

\noindent\textbf{Design Space Expansion}. Extracting design decisions from news text is challenging: open-ended prompts often lead LLMs to produce messy, inconsistent annotations, while being overly restrictive may limit their ability to uncover new design decisions.
We developed an iterative two-step extraction process to balance flexibility and consistency. 
In the first step, we extracted values for design decisions covered in our current design space, allowing us to discover new domain-specific values for existing decision types. In the second step, we prompted the LLM to exclude the key-value pairs identified in the previous step and apply the 4W1H scheme (What/When/Who/Where/How)~\cite{kim20125w1h, alhazmi2021m} to systematically uncover new decision keys. 

After each iteration, we enriched the design space by expanding value sets or adding new keys. For universal keys with domain-specific values, we incorporated new options uncovered across different domains. Additionally, we introduced new domain-specific keys and their associated values. We repeated the process until the design space reached saturation, with diminishing gains in new keys or values.

This iterative mining process not only extracts a broad set of privacy design decisions and their possible values, but also reveals how these decisions commonly appear together in real-world contexts. These co-occurrence patterns enable us to rank and prioritize design decisions during question generation (Section~\ref{sec:question_prioritization}).

\section{Question Generation Workflow}
\label{sec:question_generation}

\name leverages the underlying privacy representation to dynamically generate questions from the design space and prioritize questions based on developer’s prior responses and domain context.




\subsection{Generating Question Candidates}
\label{sec:candidate}
\name generates two types of questions to systematically explore the privacy design space: (1) \textit{exploratory} questions that add new nodes (data action or stakeholder interaction) to the data flow; (2) \textit{exploitative} questions that draw on the existing set of design decisions to elicit more detailed information about specific nodes. As developers respond, \name dynamically updates the underlying representation to reflect their answers.

\sssec{Exploratory Questions}. We prompt the LLM to propose new data actions (e.g., whether the data is stored, shared with external parties) or new stakeholder interactions for existing data action nodes (e.g., whether data collection obtains user consent, requesting deletion of stored data). We intentionally limit exploratory questions to a binary yes/no format to reduce cognitive load on developers. A ``yes'' response triggers the addition of a new node to the representation, while ``no'' responses do not make any changes. During question generation, LLM also annotates the related node in the representation and identifies potential new connections to enable automatic updates to the underlying representation.


\sssec{Exploitative Questions}. \name maintains a pool of candidate design decisions. When adding a new node to the underlying representation, \name retrieves relevant design decisions from our curated design space and enriches the pool. To ensure contextual relevance, \name annotates domain labels for the current design practice based on user input, and filters out irrelevant data practices mined in Section~\ref{sec:mine_design_space} based on domain label similarity. We use the Jaccard similarity of two domain label sets to quantify the relevance of data practices: 
$$J(S_1, S_2) = \frac{|S_1\cap S_2|}{|S_1\cup S_2|}$$
Only data practices with a similarity score above 0.4 are retained. From this filtered set, \name selects design decisions that match the type of the current node and incorporates them into the decision pool. 



Once an exploitative question is presented and answered by the user, \name prompts the LLM to generate follow-up questions to probe corner cases and contextual specifics. For example, if the system asks, ``\textit{Do you limit API usage for third-party integrations?}'', a follow-up question might be, ``\textit{What is the maximum number of API calls allowed per third party per day?}'' In doing so, \name progresses from broad privacy concepts to specific implementation details through exploration, exploitation, and follow-up questions with varying levels of granularity.






\subsection{Prioritizing Key Questions}
\label{sec:question_prioritization}

\name prioritizes which questions to ask based on co-occurrence statistics and contextual relevance to efficiently explore the privacy design space.

\sssec{Ranking Design Decisions by Co-Occurrence}. We rank design decisions using their co-occurrence frequency in the filtered corpus of relevant data practices described in Section~\ref{sec:candidate}. The mutual information (MI) score~\cite{neuman2013metaphor} between two design decisions $D_1$ and $D_2$ is calculated as:

$$MI (D_1, D_2) = \log \left(\frac{P(D_1\wedge D_2)\times N}{P(D_1)\times P(D_2)}\right)$$
where $P(D_1\wedge D_2)$ is the probability of the two decisions co-occurring, 
$P(D_1)$ and $P(D_2)$ are their individual probabilities, and $N$ is the total number of data practices involved. 

\name then uses the average co-occurrence statistics with all prior design choices to rank all design decisions in the pool. The hypothesis is that design decisions that frequently co-occur in privacy news or literature are more likely to be correlated and critical. At each step, \name generates questions for the top-$k$ decisions from the ranked list, as expanding too many decisions would significantly increase computational cost. For each prioritized decision, we provide the LLM with a key (the design decision) and its value set (possible choices), and prompt it to contextualize the question based on the developer's previous answers and the current underlying representation. To prevent biased suggestions, we prompt the LLM to present all possible design choices objectively and intentionally avoid framing that could prime developers. We used $k=3$ in our setting.








\sssec{Explore vs. Exploit}. \name maintains separate lists for exploratory and exploitative questions, and uses simple heuristics to switch between them. 
The choice depends on factors including how many questions have been answered, how the user has responded to earlier questions, and the current structure of the underlying representation. 
 As more questions are answered, the system gradually increases the probability of selecting exploitative questions to refine existing components. If a user skips an exploratory question, it interprets this as a signal that the topic may be irrelevant or already addressed, and switches to exploitative questions. The system also checks for missing node types (e.g., notice, share), and asks exploratory questions to fill those gaps.
 In addition to these automated heuristics, users can manually switch between these two modes at any time.


To avoid asking repetitive questions, \name prompts the LLM to check each new question against the question history and the current system representation. If a duplicate is detected, \name simply fetches the next candidate question.

\sssec{Termination of Question Generation}. Inspired by the Akinator game~\cite{sasson2023mirror}, \name provides three ways to end question generation. First, it imposes a hard limit of 25 questions to avoid overwhelming users. Second, it applies heuristics to stop early when enough information has been gathered or further questions seem unproductive (i.e., no new nodes are likely to be added and all potential design decisions have low co-occurrence statistics). Third, the user can end the session at any time if they feel their design is complete or no longer want to continue. When termination is triggered, \name prompts the user to either proceed with more questions to refine details or start the assessment process.





\section{Implementation}
\label{sec:implementation}

We implemented a prototype of \name, comprising a user interface and a backend LLM service. 

\noindent\textbf{User Interface}. We implemented a user interface for developers using React to streamline the process of conducting PRAM. 
Developers first provide a high-level design goal that describes the functional needs of their system (Figure~\ref{fig:design_goal}). \name then expands this into more detailed functional requirements, surfacing the privacy design decisions developers need to consider (Figure~\ref{fig:engineering_requirement}). Developers can review and modify these as needed. Here, the LLM is used to outline key data flows for initializing the privacy representation, rather than generate full system requirements.


The next stage involves a question-answering process (Figure~\ref{fig:question_answering}). The interface displays the underlying system representation at the bottom of the screen, while the top section presents questions along with relevant contextual information. The system highlights the node associated with the current question in the representation and provides a detailed view of that node next to the question to aid understanding. Users can choose to hide the visual representation and focus solely on the question view. Users may select multiple answers they believe are correct. If none apply, they can provide a custom response or skip the question. They can also return to earlier questions and revise their answers at any point.
 As users progress, the interface also provides real-time updates of the underlying representation to help users understand how their responses influence the system model. 


After answering several questions, users are prompted to either continue exploring more detailed questions or proceed to system assessment. In the assessment phase, the interface uses the LLM to generate a table summarizing data actions, types of data involved, and contextual factors (Figure~\ref{fig:system_assessment}), which users can edit directly. The LLM also suggests summary issues for each data action, which users can validate, discard, or supplement with their own insights.

Once this process is complete, users can export the results into a PRAM worksheet (Figure~\ref{fig:generated_worksheet}). We generate the worksheet using the ExcelJS library to ensure it follows the official PRAM format.

\sssec{Backend LLM Service}. The backend service handles requests from the user interface to fetch the next question. It generates questions asynchronously using the \verb|claude-3-7-sonnet-20250219| model to reduce response latency. We selected this model for its strong instruction-following capabilities, stable output quality and cost-effectiveness. For each user session, the service maintains a separate question pool and selects the next question dynamically, following the process described in Section~\ref{sec:question_generation}. To ensure question quality, we prompted the LLM to (1) ground each question in the underlying representation to ensure privacy relevance, (2) present design choices objectively to avoid judgment or biased framing (e.g., using ``would'' instead of ``should''), (3) generate questions and options that are specific, concrete, and contextualized (e.g., replacing abstract terms with clear examples) while remaining concise, and (4) reference the current representation and question history to avoid duplicates. We set \texttt{temperature} = 0, \texttt{top\_p} = 0.95 to minimize LLM output variability and maintain consistent question generation.

\section{Evaluation}

We conducted an experiment to evaluate our system's effectiveness in covering key privacy design decisions and a user study to assess the usability of our tool for developers.

\subsection{Case Studies}
\label{sec:case_study}

We evaluated \name's coverage of key privacy design decisions on real-world data practices.


\subsubsection{Dataset} We curated a dataset of 30 data practices with problematic privacy design decisions from the literature on privacy incidents~\cite{li2024redesigning, rubinstein2013privacy} and news reports of practices that triggered significant user backlash~\cite{chowdhry2016uber, carlson2010warning}. We provide detailed descriptions of the dataset in Table~\ref{tab:dataset}.

\subsubsection{Ground Truth} 
We employed a multi-stage approach~\cite{wu2022reasonable} to identify key design decisions underlying these data practices, which combined individual ideation, peer review and expert evaluation. We first conducted a structured brainstorming session~\cite{cooke1994varieties} to enumerate privacy design decisions associated with these data practices. The session involved 11 students who had extensive research training in usable privacy, including advanced coursework, active research projects, and prior publications in privacy-related venues. We provided each participant with a general description of the practices and asked them to individually propose design decisions they felt significantly affected user privacy. They then worked in pairs to discuss their responses and identify additional design decisions.

The identification phase was followed by a ranking and selection stage conducted by more experienced researchers to identify the key design decisions. After the session, two graduate researchers with over three years of experience in usable privacy research independently coded the design decisions for each data practice based on participant input.
We assessed inter-rater reliability using Cohen’s kappa~\cite{mchugh2012interrater}, which indicated strong agreement with $\kappa = 0.85$.
The coders then discussed any discrepancies to resolve conflicts and reach a consensus on the codebook. After coding, they ranked the decisions by their frequency of occurrence across participants and perceived impact on user privacy, then selected the top-ranked decisions to serve as the ground truth.



\subsubsection{Method}
We manually defined a design goal for each data practice based on its functionality (e.g., “an attendee tracking feature for a video conferencing app” for Zoom's attention tracking), which served as the input to our system. We then answered the questions generated by \name to align the responses with real-world implementations.
For each data practice, we measured whether the resulting privacy representation could (1) capture design decisions by including relevant keys, and (2) accurately reflect actual choices. We used the GPT-4o model to compare the generated outputs against our ground truth, and supplemented this with human review and corrections to ensure the accuracy of evaluation.  
We also tested whether the effectiveness in capturing privacy-related design decisions primarily stems from our question-generation workflow or LLMs’ inherent capabilities. For baselines, we directly prompted different LLMs to generate 20 questions. We set the following configurations for all models: \texttt{temperature} = 0, \texttt{top\_p} = 0.95, and \texttt{max\_tokens} = 4000. We tested each model three times to account for randomness and averaged the results.




\subsubsection{Results} Table~\ref{tab:benchmark} compares the coverage of key privacy design decisions of \name with three state-of-the-art LLMs, including GPT-4o, Claude 3.7, and Gemini 2.5 Flash.
On average, \name achieved a coverage rate of 93.67\% for key decisions and 77.33\% for actual choices, significantly outperforming the baseline LLM approaches, whose coverage ranged from 43.91\% to 55.22\% for key decisions and 20.83\% to 42.71\% for actual choices.

We further evaluated their performance across 10 categories of privacy decisions (4 data actions \& 6 stakeholder interactions). We found that \name consistently showed higher coverage on key design decisions and their actual choices across all categories. The performance advantages were more significant in data action categories (i.e., Collect, Process, Store, and Share), where \name achieved coverage rates above 89\% for key decisions, while baseline LLMs typically ranged between 38--65\%.


Stakeholder interaction categories revealed more nuanced performance differences. While LLMs performed relatively well in the Consent category (75.72\% for GPT-4o, 78.44\% for Gemini 2.5), \name still led with coverage of 96.15\% for key decisions and 88.46\% for actual choices. The most significant gaps appeared in Control and Access categories, where baseline LLMs struggled to identify relevant design decisions (40.65--62.16\% coverage), while  \name maintained over 92\% coverage.

These results suggest that \name is more effective at systematically exploring the privacy design space. While general-purpose LLMs can generate contextual privacy-related questions, their outputs are often inconsistent and less effective due to limited domain-specific knowledge.




\begin{table*}[t]
    \centering\small
\caption{\name achieved broader coverage of key privacy design decisions and their corresponding values than three baseline models (i.e., GPT-4o, Claude 3.7, and Gemini 2.5 Flash) across all categories.}
\Description{A table comparing the performance of PrivacyAkinator against three baseline LLM models (GPT-4o, Claude-3.7, Gemini 2.5 Flash). For 10 categories of privacy decisions (e.g., Collect, Consent), the table shows the percentage of key Decisions and actual Choices that each model successfully identified.
}

    \begin{tabular}{c cc cc cc cc}
    \toprule
     \multirow{2}{*}{Coverage}  & \multicolumn{2}{c}{GPT-4o} & \multicolumn{2}{c}{Claude-3.7} & \multicolumn{2}{c}{Gemini-2.5 Flash} & \multicolumn{2}{c}{\name}
     \\\cmidrule(lr){2-3}\cmidrule(lr){4-5}\cmidrule(lr){6-7}\cmidrule(lr){8-9}
     & Decision & Choice & Decision & Choice & Decision & Choice & Decision & Choice \\
    \midrule
    Collect & 59.74 \% & 29.47 \% & 60.90 \%  & 30.77 \% & 65.38 \% & 55.13 \% & \textbf{98.08 \%}  & \textbf{80.77 \%}  \\
    Process & 38.38 \% & 18.84 \% & 55.10 \%  & 28.57 \% & 54.42 \% & 40.14 \% &  \textbf{89.80 \%}  & \textbf{75.51 \%} \\
    Store & 55.64 \% & 35.16 \% & 55.26 \% & 30.70 \% & 75.27 \% & 56.76 \% &   \textbf{100.00 \%} & \textbf{92.11 \%} \\
    Share & 43.54 \% & 18.62 \% & 42.67 \% & 22.67 \% & 48.50 \% & 36.44 \% &  \textbf{92.00 \%} & \textbf{80.00 \%} \\
    \midrule
    Consent & 75.72 \% & 44.06 \% & 63.28 \% & 41.13 \% & 78.44 \% & 62.28 \% & \textbf{96.15 \%} &  \textbf{88.46 \%} \\
    Notice & 54.01 \% & 26.41 \% & 68.69 \% & 32.28 \% & 60.00 \% & 44.78 \% &  \textbf{89.47 \%} & \textbf{63.16 \%} \\
    Control & 40.65 \% & 15.34 \% & 62.16 \% & 31.96 \% & 59.74 \% & 54.62 \% & \textbf{92.00 \%} & \textbf{64.00 \%} \\
    Access & 45.68 \% & 27.16 \% & 53.53 \% & 22.13 \% & 60.49 \% & 48.15 \% &  \textbf{92.59 \%} &  \textbf{81.48 \%} \\
    Request & 29.63 \% & 7.41 \% & 40.74 \% & 22.22 \% & 40.74 \% & 25.93 \% &  \textbf{88.89 \%} & \textbf{66.67 \%} \\
    Audit & 40.00 \% & 6.67 \% & 94.44 \% & 48.41 \% & 53.33 \% & 40.00 \% &  \textbf{100.00} \% &  \textbf{60.00 \%} \\
    \midrule
    Overall & 43.91 \% & 20.83 \% & 54.25 \% & 28.26 \% & 55.22 \% & 42.71 \% &  \textbf{93.67 \%} &  \textbf{77.33 \%} \\
    \hline
    \end{tabular}
    \label{tab:benchmark}
\end{table*}

\subsection{User Studies}
\label{sec:user_study}


We conducted a user study to evaluate how \name could help developers articulate privacy design decisions.


\subsubsection{Participants}
We recruited 24 participants (16 identified as male and 8 as female; 18 aged 18--24, 6 aged 25--34) in the evaluation through the same approach as previous studies (Section~\ref{sec:study_method}). 
Each participant received a USD 20 Amazon gift card as compensation. Participants reported low levels of prior privacy knowledge (Mean = 2.4, SD = 1.3), with most rating themselves as beginners (ratings of 1--2). None of the participants had taken part in our earlier studies. The study received approval from our university’s Institutional Review Board (IRB).


\subsubsection{Study Procedure and Apparatus}
The study used a within-subjects design, where participants used both \name and NIST's PRAM to complete two tasks. Because this study focuses on novices, we anticipated that the individual differences in the between-subjects design could strongly influence outcomes. We randomized the presentation order of tools and tasks, and included warm-up tasks for each tool to mitigate the potential priming effect. We did not disclose that one of the tools was developed by the authors
to prevent introducing potential bias. For PRAM, we only used \textit{Worksheet 2 (Assessing System Design)} to ensure fair comparison, as it aligns with the focus of \name on articulating system design decisions.
We reused the task scenarios from previous studies (Table~\ref{tab:nist_study}).


Each study included two sessions, one for each condition. Each session started with a 10-minute walk-through briefing on the study purpose and task overview, followed by a warm-up task that included a tutorial on the tool and the task using a sample scenario \textit{ACME IDP service} drawn from the PRAM materials.  Researchers observed the process and answered questions as needed. 

Then we instructed each participant to complete one task using their assigned tool (PRAM or \name) to assess the privacy design of one data practice scenario. We presented a high-level description of each scenario (Table~\ref{tab:nist_study}) to remain consistent with previous studies (Section~\ref{sec:nist_study}). 

 Upon completion of each task, we asked the participant to fill out a NASA-TLX survey~\cite{hart1988development}. After both sessions, participants took a semi-structured interview about their experiences with the two tools. In the post-study interview, we asked participants to compare their experiences with the two approaches and reflect on the strengths and weaknesses of each. The study was conducted in a lab space or via an online Zoom meeting. On average, our study took about 75 minutes for each participant. 

\subsubsection{Data Analysis}
We measured the coverage of key privacy design decisions identified in participants’ responses. Two researchers collaboratively annotated the design decisions described in participants’ worksheets (written manually or generated by \name), counted the \textbf{total number} of decisions included in each worksheet, and calculated the \textbf{coverage} of each design decision by comparing them to the ground truth for each scenario (Section~\ref{sec:case_study}). During annotation, we excluded content that did not represent design decisions, such as user privacy expectations (e.g., ``\textit{this data is highly sensitive for users}''), and removed duplicate entries. We performed Mann-Whitney U tests~\cite{mcknight2010mann} to compare different scenario orders within each condition and found no evidence of ordering bias across all reported quantitative metrics. The qualitative analysis followed the same procedure as described in Section~\ref{sec:study_method}.

\begin{table*}[htbp]
    \centering\small
\caption{In our user study, participants spent less time and covered more key design decisions with \name compared to PRAM Worksheet. Format: mean $\pm$ standard deviation. We also highlight the higher value between the two conditions.}
\Description{A table showing quantitative results from a user study. For four different tasks, it compares the performance of participants using the standard PRAM worksheet versus \name on three metrics: Time (min), # Total Decisions identified, and Key Decision Coverage percentage. A final row provides the average for each metric.
}
    \begin{tabular}{c c c c c c c}
    \toprule
    \multirow{2}{*}{Task ID} & \multicolumn{3}{c}{PRAM Worksheet} & \multicolumn{3}{c}{\name}
    \\\cmidrule(lr){2-4}\cmidrule(lr){5-7}
    & Time (min) & \# Total Decisions &  Key Decision Coverage & Time (min) & \# Total Decisions & Key Decision Coverage \\
    \midrule
    Task \#1 & 36.4 $\pm$ 10.9 & 12.4 $\pm$ 3.5 & 36.9 $\pm$ 13.8 \%  & \textbf{14.2} $\pm$ 9.6 & 33.5 $\pm$ 5.9 & \textbf{79.1} $\pm$ 3.8 \% \\
    Task \#2 & 32.6 $\pm$ 11.1 & 11.2 $\pm$ 3.6 & 40.0 $\pm$ 16.9 \%  & \textbf{7.0} $\pm$ 1.9 & 29.8 $\pm$ 6.2  & \textbf{93.3} $\pm$ 6.2 \% \\
    Task \#3 & 31.5 $\pm$ 5.2 & 11.4 $\pm$ 3.2 & 38.2 $\pm$ 7.6 \% & \textbf{7.5} $\pm$ 1.8 & 26.3 $\pm$ 4.3  & \textbf{93.9} $\pm$ 4.1 \% \\
    Task \#4 & 41.8 $\pm$ 9.3 & 13.3 $\pm$ 6.0 & 48.6 $\pm$ 9.8 \%  & \textbf{11.2} $\pm$ 2.8  & 24.8 $\pm$ 1.7 & \textbf{87.5} $\pm$ 4.8 \%  \\
    \midrule
    Average & 35.8 $\pm$ 10.7 & 12.9 $\pm$ 4.3 & 42.6 $\pm$ 12.0 \%  & \textbf{9.7} $\pm$ 6.4 & 28.9 $\pm$ 5.5  & \textbf{89.3} $\pm$ 7.0 \%  \\
    \bottomrule
    \end{tabular}
    \label{tab:developer_study}
\end{table*}

\subsubsection{Quantitative Results} 
We found that \name could help developers articulate system design more effectively and improve the quality of their design outputs.


\sssec{Improved Efficiency of Articulating System Design}.  As shown in Table~\ref{tab:developer_study}, participants completed the task 73\% faster (average of 9.7 vs. 35.8 minutes) and covered 47\% more privacy-related design decisions (average of 28.9 vs. 12.9) using \name compared to the PRAM worksheet.

Additionally, participants reported more positive responses to the NASA TLX questions (Table~\ref{tab:tlx}) with \name than the PRAM worksheet. These improvements include lower mental workload, effort, and frustration, along with increased self-perceived performance, which are all statistically significant under the Wilcoxon Signed Rank test~\cite{woolson2005wilcoxon} ($p < 0.01, r > 0.8$).

\sssec{Broader Coverage of Key Privacy Design Decisions}. We then compared the coverage of design decisions between conditions (Table~\ref{tab:developer_study}). On average, participants covered 89.3\% of key privacy design decisions when using \name, compared to 42.6\% of the PRAM worksheet. This difference was statistically significant under the Wilcoxon Signed Rank test~\cite{woolson2005wilcoxon} ($p < 0.01$).  Additionally, we observed that \name produced more consistent results across participants, with lower variance in coverage (7.0\%) than 12.0\% for the PRAM worksheet.

\begin{table*}[t]
\small
\caption{Participants perceived significantly lower cognitive load using \name compared to PRAM worksheet. We present NASA TLX results (scale of 1 to 7) as ``median (mean
$\pm$ standard deviation)''. We annotate statistical significance based on the Wilcoxon Signed-Rank test $(^*: p < 0.05,  ^{**}: p < 0.01,  ^{***}: p < 0.001)$. $\downarrow$ denotes that the lower is better.}
\Description{A table presenting results from the NASA Task Load Index (TLX) survey, comparing the perceived cognitive load of using PRAM versus PrivacyAkinator. The columns represent six dimensions of workload: Mental, Physical, Temporal, Performance, Effort, and Frustration. Arrows indicate whether a lower or higher score is better. Asterisks denote statistically significant differences.
}
\begin{tabular}{c cccccc}
\toprule
 Condition  & Mental~$\downarrow$ & Physical~$\downarrow$ & Temporal~$\downarrow$ & Performance~$\uparrow$ & Effort~$\downarrow$& Frustration~$\downarrow$\\
\midrule
PRAM & 6 (5.4 $\pm$ 1.4)  & 2 (2.5 $\pm$ 1.8) & 3 (3.6 $\pm$ 1.6) & 5 (4.3 $\pm$ 1.1) & 5 (5.3 $\pm$ 1.0) & 3 (3.4 $\pm$ 2.0) \\
\name & \textbf{3 (3.3 $\pm$ 1.3) ***} & \textbf{1 (1.7 $\pm$ 1.1) *} & \textbf{2 (2.1 $\pm$ 0.9) ***} & \textbf{6 (5.7 $\pm$ 0.6) ***} & \textbf{3 (3.1 $\pm$ 1.5) ***} &  \textbf{1 (1.7 $\pm$ 1.0) **} \\
\bottomrule
\end{tabular}
\label{tab:tlx}
\end{table*}

\subsubsection{Qualitative findings}
We made the following findings based on participants’ task behaviors and post-study interview feedback.

\sssec{Lower Barriers for Privacy Non-Experts}. 
Participants appreciated \name's guided question answering process, which enabled them to describe privacy-related decisions without specialized privacy background. For example, P16 mentioned that ``\textit{[PrivacyAkinator] is pretty intuitive and easy to use. It doesn't require any extra privacy knowledge.}'' P18 added that ``\textit{Personally, I haven't seen that workflow before, and I haven't systematically studied privacy. But I could still get where each design was coming from and the kind of details they were trying to address. The questions made a lot of sense to me, and they also gave me a few solid options to think through}.''

\sssec{Ease of Cognitive Load}. Participants expressed that \name streamlined the system design articulation process, reducing their perceived effort and workload. 
Many participants described \name as ``\textit{much easier to use}'', with P12 noting that ``\textit{the only mental effort was answering the questions which mainly had just 2 or 3 options}'', rather than ``\textit{having to come up with a lot of ideas and also decide on which idea to focus on}.'' P4 mentioned that ``\textit{[PrivacyAkinator] was much more efficient for my thought process, and it allowed me to generate more ideas and also not focus about manually drawing the diagram}.'' Participants also recognized that \name ``\textit{made the specification process easier and more manageable}'' by ``\textit{narrowing the mental search space}'' (P18), and offering a simplified workflow that avoided ``\textit{doing all things manually}'' (P4)  and ``\textit{switching back and forth between different tabs}'' (P15).






 

\sssec{Enhanced Coverage of Privacy Design Decisions}. 
Participants appreciated that \name helped surface design considerations they might have otherwise overlooked.
P13 shared that ``\textit{questions can help me to find those details I missed, or even just didn't know}'', while P1 noted that ``\textit{the set of questions presented gave a lot of insights into designing and addressing data privacy issues in the real world projects or applications}.''
Participants also perceived better performance with \name than designing from scratch. P11 explained that without guidance, they were ``\textit{just imagining things from scratch},'' and felt ``\textit{limited in what I was able to come up with}.'' In contrast, \name ``\textit{gives something to build on and [they] just needed to make some reasonable adjustments}'', resulting in a design that they felt was more accurate and complete.




\sssec{Reduced Ambiguity through Structured Guidance}.
Participants expressed that \name reduced their confusion by providing clear, step-by-step guidance throughout the process. Most participants struggled with traditional worksheets. They explained that ``\textit{the worksheet definitions and examples are vague and sometimes disagreeing}'' (P7), and initially felt ``\textit{kind of lost}'' and ``\textit{having no idea what you're gonna do}'' (P19). In contrast, Participants appreciated how \name used contextualized prompts, a visual workflow diagram, and follow-up questions to make the process ``\textit{easier to understand}''. P1 appreciated that ``\textit{[PrivacyAkinator] infers based on the questions answered, which serves as a great way of summarizing and visualizing details while designing a system.}'' P6 also noted that ``\textit{questions generated by \name were easier to understand,}'' because they avoided heavy use of technical jargon.


\sssec{Over-Reliance on Generated Design Decisions}.
Participants also noted the potential downside of over-relying on system-generated questions. P11 observed that when
answering the generated questions, they are ``\textit{kind of shut off the thinking part of your brain, and then don't explore beyond it}.''
While participants recognized that \name is especially useful for beginners with limited knowledge, they suggested a hybrid approach for experienced developers. As P5 explained, ``\textit{If you're more experienced, the tool is still good, because it'll give you different options. But you should still take time to sit down and come up with your own ideas, instead of just relying on what it's generating.}''

\section{Discussion \& Limitations}
\label{sec:discussion}


\sssec{New Design Knowledge}. Our studies suggest several design implications for developing tools that support privacy assessment. Existing privacy assessment frameworks often prompt developers to analyze risks before they have fully articulated the underlying data practices~\cite{enterprise2020nist, wuyts2020linddun}. Our studies show that this leads to incomplete or inconsistent system descriptions. Without structured articulation, novices rely on ad-hoc reasoning and tend to fixate on the first thing they notice, resulting in partial and biased assessments. Our findings suggest that privacy assessment tools should defer judgment and \textbf{separate articulation from evaluation} [\textit{Design Implication} \#1]. Articulating data practices before assessing risks supports a more complete understanding of the system and reduces premature evaluation.


\textbf{RQ1: How to turn privacy design into a structured task?}
 We address this through a structured representation that organizes privacy design decisions around data flows and stakeholder interactions. In our studies, novices were often distracted by surface-level implementation details, echoing prior findings that novices tend to focus on superficial details whereas experts attend to deeper structural cues~\cite{ngoon2018interactive, burkhardt2002object, bransford2000experts}. Our representation helps bridge this gap through \textbf{selective attention} [\textit{Design Implication} \#2]: by determining what to include and what is omitted, it not only documents the system design, but also actively directs developers' attention and guides their thinking ~\cite{norman2014things}. 

\textbf{RQ2: How to unfold this structured task for novice developers?} \name \textbf{decomposes} PRAM's broad, open-ended prompts into \textbf{a series of multiple-choice questions} [\textit{Design Implication} \#3] to make privacy assessment more approachable and actionable for novices. In our studies, participants often did not know how to begin with open-ended worksheets, felt confused about what level of detail was expected and frequently omitted key privacy decisions. With open-ended worksheet, novices need to consider numerous decisions simultaneously, which imposes high cognitive load and can be overwhelming. In contrast, our evaluation shows that \name's structured prompts improved coverage of privacy-relevant decisions and reduced cognitive load, consistent with prior work on scaffolding novices~\cite{van2024cognitive, ge2005scaffolding, stender2015scaffolding, macneil2021framing}. By breaking the task into discrete, concrete questions, the tool allows novices to focus on one decision at a time. 

A key design trade-off is reducing cognitive load without oversimplifying the task or limiting expressiveness. To address this, we decouple \textit{what} to articulate (RQ1) from \textit{how} to support articulation (RQ2): The structured representation directs attention to privacy-relevant aspects of a system, while the interactive questions guide novices through the design space. Our design implications (e.g., separating articulation from evaluation, guiding selective attention, and decomposing complex tasks) can generalize beyond \name and inform the design of related systems. Designers can operationalize these principles in domain-specific contexts; for example, we use closed-ended questions to scaffold privacy risk assessment tasks.

\sssec{Practical Deployment Limitations}.
While \name helps developers articulate their design choices, privacy risk assessment involves additional challenges beyond the tool’s current scope. Improved coverage and articulation may not guarantee correct privacy decisions. Judging the severity of privacy risks remains subjective and often requires domain-specific expertise that novices may lack. 
Nevertheless, by constructing a closed-ended, structured design space, our work lays the groundwork for creating a representation of privacy design that is easier to analyze with automation. 
Future work could build on this representation to produce machine-readable outputs that support risk identification and prioritization—for example, by training ML models to classify risk levels, flag high-risk combinations of design choices, or map design patterns to known risks—thereby enabling more objective and consistent privacy assessments.

%


One related concern is the potential over-reliance on automated suggestions. \name is not intended to replace existing PRA workflows but to serve as a structured prompt or checklist to help developers surface key design decisions. 
Developers should combine the automated question generation with manual review and domain expertise, as the LLM-based approach may miss context-specific privacy considerations.



\noindent\textbf{Potential Bias of Mining Privacy News}.
As with any data-driven application, the coverage of \name is influenced by its data sources. Since we construct the privacy design space by mining privacy-related news articles, it effectively captures high-profile, media-reported incidents but may overlook under-reported or emerging domains, such as healthcare IoT or industrial surveillance. For example, it may over-represent issues that were widely reported rather than those that are actually more common in practice. Additionally, the news corpus may reflect geographic bias, as the sources are drawn primarily from media outlets in the United States and United Kingdom, which could underrepresent privacy issues more prominent in other regions or regulatory environments.

Additional sources (e.g., legal documents) can be incorporated to broaden the system’s coverage and relevance to sensitive contexts. Organizations can also apply our approach to create customized taxonomy using internal or domain-specific materials, such as engineering requirement documents.



\sssec{Risks of Relying on LLM-generated Content}. 
\name relies on LLMs to generate questions, which can sometimes result in vague or inconsistent wording. During our user study (Section~\ref{sec:user_study}), several participants noted that some questions were repetitive or unclear. To mitigate this, \name allows users to skip ambiguous or duplicate questions and submit custom responses when the provided options do not apply. Repetitive questions may cause inconsistencies. Future work could analyze the representation and provide automatic or human-assisted conflict resolutions. 
For vague or unclear questions, a conversational assistant could be developed to provide explanations, or regenerate questions for clarity.
With the recent rapid advancements in LLMs~\cite{yi2024survey, chang2024survey}, we expect future models to reduce these issues and further improve the clarity and quality of generated questions.

Unconstrained question generation carries a high risk of LLM hallucinations, potentially producing nonsensical or irrelevant questions that do not connect to the developer’s system. Instead of allowing free-form brainstorming, we ground the questions on the closed-ended design space and provide contextual cues (e.g., highlighting related nodes, summarizing prior decisions) to help developers flag irrelevant questions.


Another concern is that LLMs' output is not necessarily deterministic and may vary across runs. In our user studies, however, we observed high question stability: repeated invocations for same system descriptions produced largely consistent question sets, with variation mostly confined to minor wording changes. We attribute this stability to low-temperature sampling and the system’s prioritization logic, which limits randomness in question selection. Future work may consider alternative question-generation approaches (e.g., template-based methods), which offer higher stability but may lack the flexibility to capture diverse, context-specific privacy nuances. For example, it would be challenging for templates to surface questions tailored to a developer’s specific data uses.

Sharing system descriptions with an external LLM may expose proprietary system logic and sensitive data-processing details. Developers can address these privacy risks by establishing contractual agreements with commercial LLM providers (e.g., zero data retention policies~\cite{gupta2025zerodataretentionllmbased}) or by deploying self-hosted models.

LLM interpretations of system descriptions may be inaccurate, leading to misrepresented features, data uses, or processing steps. To help detect these misinterpretations, \name immediately visualizes how each input updates the underlying representation, making the LLM’s interpretation explicit and easier to review.

\name uses an LLM to generate preliminary system requirements from a short description. Recent work in requirements engineering shows that modern LLMs (e.g., GPT-4) can reliably interpret natural-language descriptions and produce structured, coherent software requirements specifications comparable to those written by entry-level engineers~\cite{krishna2024using, hymel2025analysisllmsvshuman, ebrahim2025enhancing}. For example, in the Zoom-attention scenario, the model recognizes that application-window focus and webcam-based inference are two distinct methods for tracking attention. Its output uses window focus as it is less privacy-invasive, given that much sensitive information beyond attention status can be inferred from webcam data (e.g., facial expressions, body movements).
Alternative methods for requirement elicitation, such as a drag-and-drop interface, offer more direct control over feature and data selection but may lack generalizability. For example, a predefined list of data types may struggle to capture the breadth of real-world design choices. Such approaches are better suited when customized for a specific organization or domain, where relevant data types are fixed and well-defined.



\noindent\textbf{Applicability beyond Novices}. 
We intentionally focused our studies on developers who are novices in privacy. Prior work shows that experts and novices often exhibit markedly different usage patterns when engaging in complex design and analysis tasks~\cite{dou2009comparing, ahmed2003understanding}, so we expect privacy experts to have different user needs that require different forms of support. Focusing on a specific user group allows us to probe their specific challenges in depth and provide tailored scaffolding, rather than a one-size-fits-all system that serves neither group well. We target novices in particular because they constitute a large portion of the overall developer population, and their lack of privacy expertise necessitates additional support for effective privacy assessments.

While \name is specifically designed for novices, we believe several of its components can also benefit privacy experts. For example, the structured privacy representation can help them keep track of numerous privacy-relevant decisions and more easily navigate the design space. The prompting questions can function as a lightweight checklist to surface missing decisions or reveal edge cases, or as prompts to support brainstorming of alternative design choices.
However, experts may demand different workflows: they often do not need step-by-step guidance and instead may prefer to manually assemble the diagrams from the building blocks (i.e., data actions and interactions), similar to privacy storyboards~\cite{jin2021lean}. They can then actively drive question generation, instructing the system to generate questions when needed to enrich the representation more deliberately. Because our studies primarily involved novices, our results may not fully reflect how experts would use or benefit from the system. Future work will test the tool with developers of varied expertise levels to incorporate expert perspectives for a more comprehensive assessment.

\section{Conclusion}
This paper explores methods to lower the barrier for novice developers to apply privacy risk assessments, particularly NIST's PRAM. Through an observational study (N=12), we identified three key challenges novice developers face when applying PRAM: (1) difficulty effectively distributing attention across numerous privacy design decisions, (2) confusion stemming from PRAM's vague terminology and open-ended structure regarding appropriate responses, and (3) frequently missing important design decisions without explicit prompting. To address these challenges, we developed \name, a tool that helps developers articulate key privacy decisions by answering LLM-generated multiple-choice
questions. Our user study (N=24) shows that \name enables developers to identify 47\% more key design decisions in 73\% less time compared to using the PRAM worksheet.

\bibliographystyle{ACM-Reference-Format}
\bibliography{sample-base}

\onecolumn

\appendix
\section{User Interface}
\begin{figure}[h]
    \centering
    \includegraphics[width=0.8\linewidth]{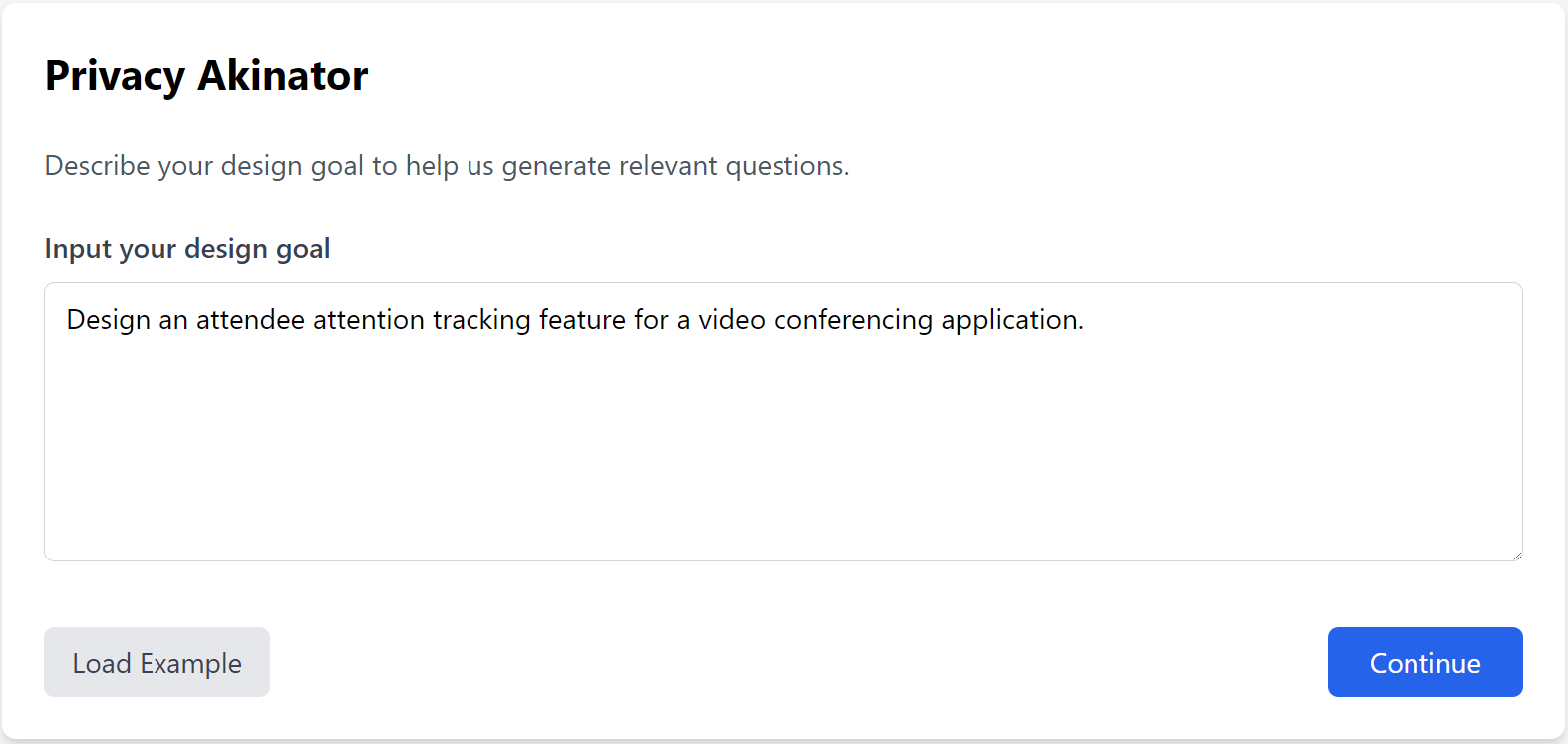}
    \caption{\name Design Goal Panel}
    \label{fig:design_goal}
    \Description{A screenshot of the initial user interface for \name. It features a title, "\name," and a large text input box labeled "Input your design goal." The box contains example text: "Design an attendee attention tracking feature for a video conferencing application." There are buttons for "Load Example" and "Continue."}
\end{figure}

\begin{figure}[h]
    \centering
    \includegraphics[width=0.8\linewidth]{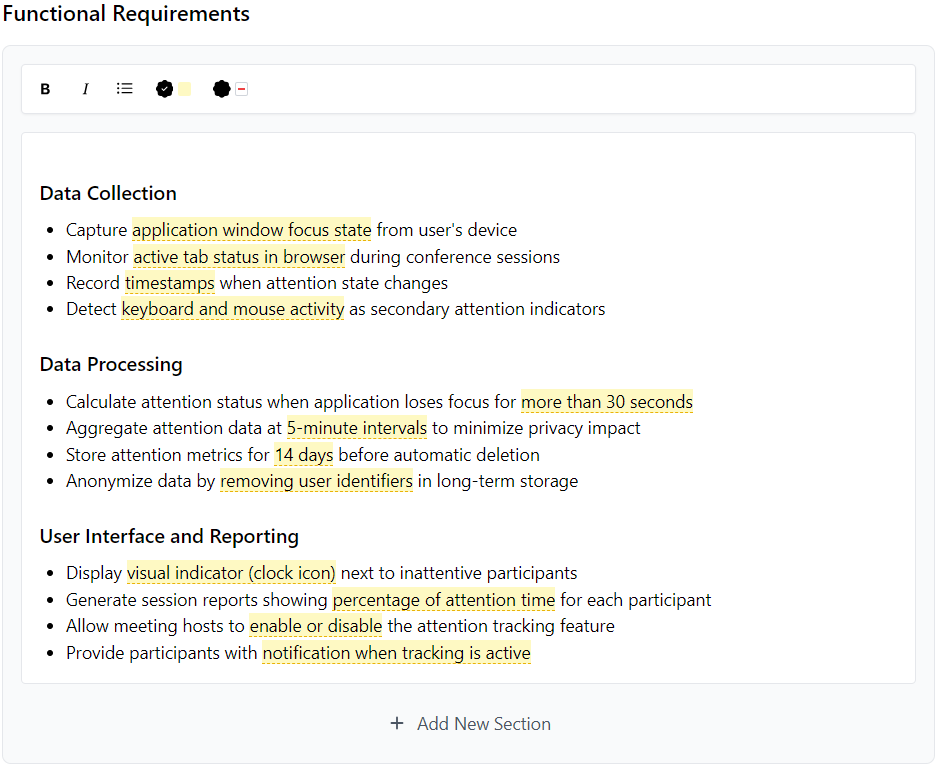}
    \caption{\name functional Requirement Panel}
    \label{fig:engineering_requirement}
    \Description{A screenshot of the functional requirements editor in \name. It shows a list of requirements organized under headings like "Data Collection" and "Data Processing." Specific requirements, such as "Calculate attention status when application loses focus for more than 30 seconds," are highlighted, indicating that these are editable design choices.}
\end{figure}

\vspace{30pt}

\begin{figure}[h]
    \centering
    \includegraphics[width=0.9\linewidth]{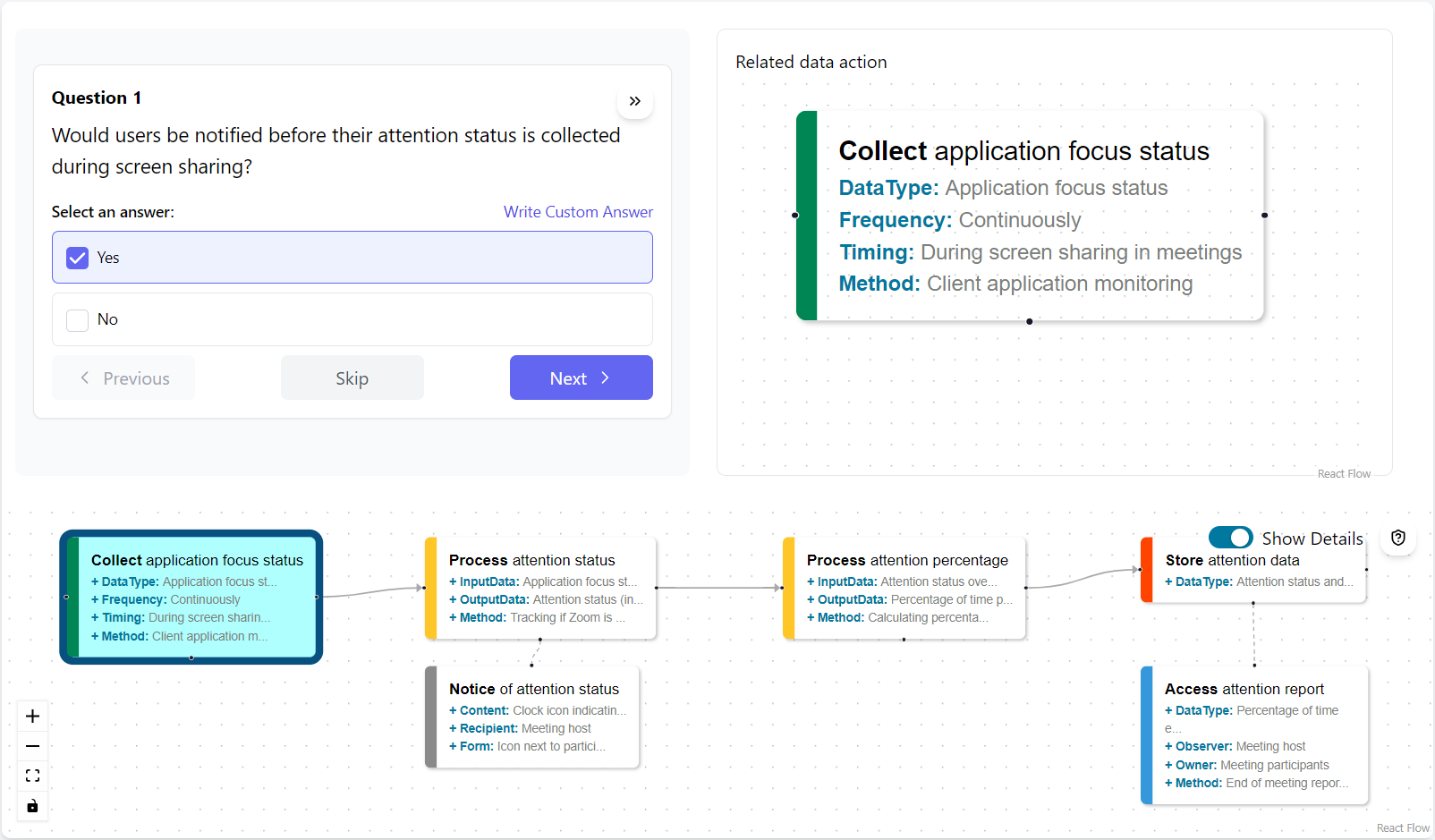}
    \caption{\name Workflow \& Questions Panel}
    \label{fig:question_answering}
    \Description{A composite screenshot showing the main interaction panel of \name. The top part displays a question for the user ("Would users be notified...?") with multiple-choice answers. To the right is the "Related data action," which provides context. The bottom part shows the multi-layer graphical representation of the system design, which updates as the user answers questions.}
\end{figure}


\begin{figure}[h]
    \centering
    \includegraphics[width=0.9\linewidth]{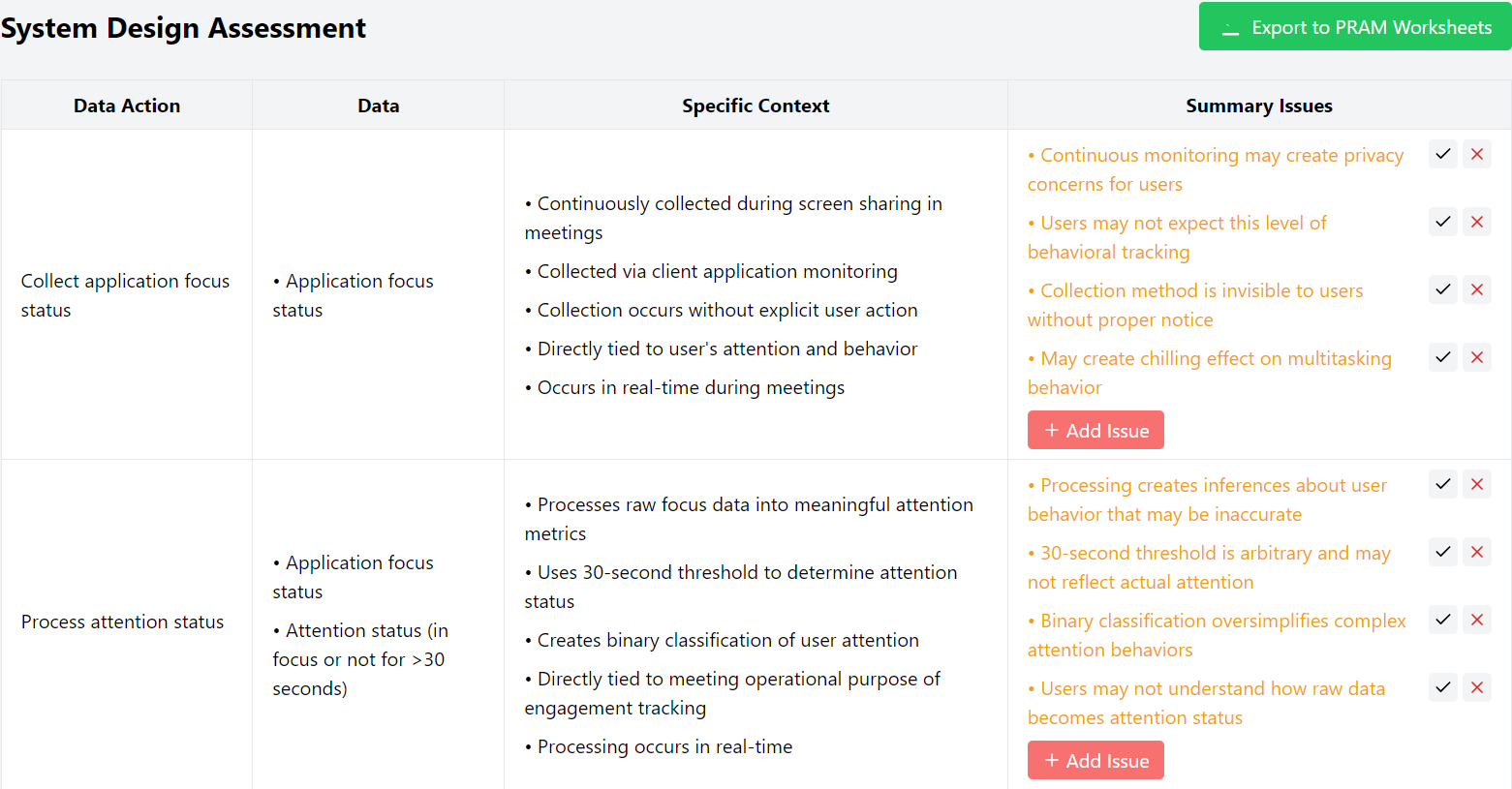}
    \caption{\name Access System Design Panel}
    \label{fig:system_assessment}
    \Description{A screenshot of the system design assessment table in \name. The table has columns for Data Action, Data, Specific Context, and Summary Issues. Rows detail specific actions like "Collect application focus status" and list potential issues, such as "Continuous monitoring may create privacy concerns for users," with options for the user to validate or edit these issues.}
\end{figure}

\clearpage
\begin{figure}[h]
    \centering
    \includegraphics[width=0.8\linewidth]{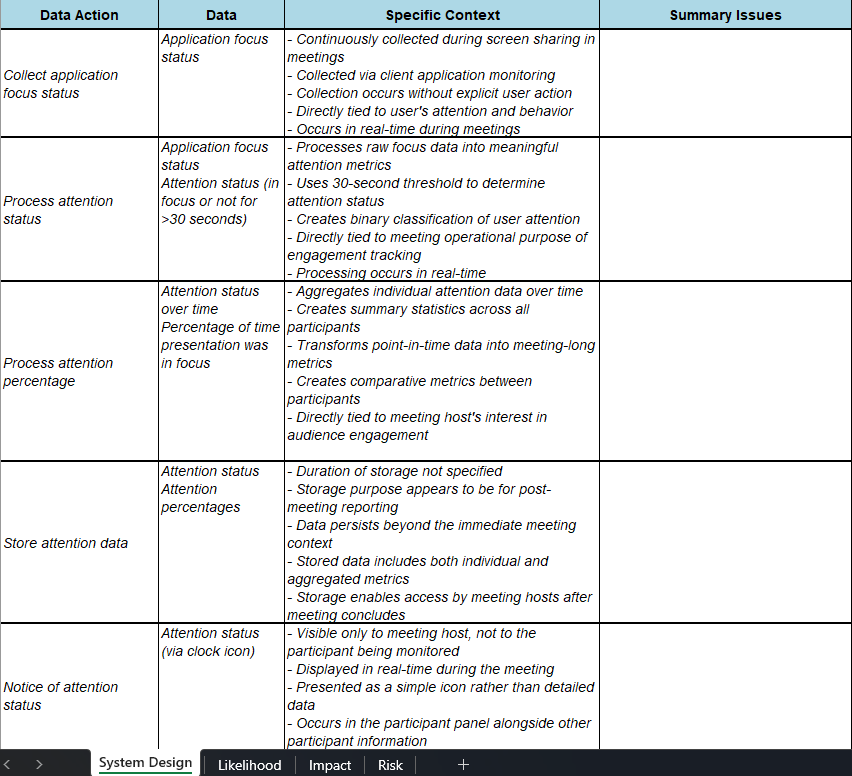}
    \caption{\name Generated Worksheet}
    \label{fig:generated_worksheet}
    \Description{A screenshot of a worksheet exported from \name, formatted for NIST PRAM. It's a detailed table with columns for Data Action, Data, Specific Context, and Summary Issues, summarizing the privacy analysis of the user's design. This shows the final output of the tool.}
\end{figure}

\section{Data Actions and Stakeholder Interactions}
\begin{table*}[htbp]
    \centering\small
\caption{Data Actions and Stakeholder Interactions}
\Description{A table defining the common operations used in PrivacyAkinator's representation. The table has three columns: Category (Data Action or Stakeholder Interaction), Operation (e.g., Collect, Consent, Control), and Definition (a brief explanation of the operation).
}
    \begin{tabular}{lll}
    \toprule
       \textbf{Category}  &  \textbf{Operation} & \textbf{Definition} \\
    \midrule
    \multirow{4}{15mm}{Data Action} & Collect & Collect users' data or sensor inputs.\\
     & Process & Process data to derive new information. \\
     & Store & Keep data in a persistent storage system. \\
     & Share & Share data to different parties. \\
    \midrule
    \multirow{7}{15mm}{Stakeholder Interaction} & Consent & A data subject gives permission for specific data actions.\\
     & Notice & A data observer informs data subjects about the data action or its results. \\
     & Control & A data subject controls settings that determine how their data is stored or processed. \\
     & Access & A data observer accesses and uses user data or derived data for a specific purpose. \\\
     & Request & A data subject asks to exercise their data rights. \\
     & Audit & An auditor examines data actions for compliance with policies or regulations. \\
     & Influence & A data beneficiary/victim is impacted by a data practice. \\
    \bottomrule
    \end{tabular}
    \label{tab:definition}
\end{table*}

\section{Selected Data Practices for Evaluation}

\begin{table*}[h]
    \centering\small
\caption{A summary of 30 data practices collected to assess the coverage of key privacy design decisions.}
\Description{A table listing 30 real-world data practices used to evaluate the coverage of PrivacyAkinator's design space. The columns are ID and Scenario (e.g., "Zoom Attention Tracking," "Target Pregnancy Prediction," "Social Credit Score"), with a brief Description for each.
}
    \begin{tabular}{p{3mm} p{43mm} p{123mm}}
    \toprule
        \textbf{\#ID} & \textbf{Scenario} & \textbf{Description} \\
    \midrule
        \#1 & Zoom Attention Tracking & An attendee attention tracking feature for a video conferencing application \\
         \hline
        \#2 & Facebook Cambridge Analytica & A data platform that allows third-party apps to collect users’ data through a social media platform \\
         \hline
        \#3 & AirTag & A real-time location tracking device for personal items \\
         \hline
        \#4 & Target Pregnancy Prediction & A retail analytics system that processes customer purchase histories to automatically generate personalized coupons and recommendations \\
         \hline
        \#5 & Facebook Emotional Contagion & A research experiment to study how exposure to emotional content affects users' own emotional expressions on a social media platform \\
         \hline
        \#6 & Google Buzz & A system that integrates email services with social networking functionality \\
         \hline
        \#7 & OKCupid Score manipulation & A research experiment to study how displayed compatibility scores influence user behavior and interactions on a dating platform \\
         \hline
        \#8 & Uber Price Discrimination & A dynamic pricing system for a ride-sharing service \\
         \hline
        \#9 & Staple Price Discrimination & An e-commerce pricing system that incorporates geographic and market data \\
         \hline
        \#10 & Expedia Price Discrimination & A dynamic pricing system for a travel booking platform\\
         \hline
        \#11 & Strava Fitbit Heatmap & A global visualization of fitness tracking data that unintentionally reveals military personnel's locations and movements in sensitive areas\\
         \hline
        \#12 & Alexa Smart Home & A voice-activated smart home assistant that minimizes data collection while maintaining functionality and user convenience\\
         \hline
        \#13 & Tesla Camera & A vehicle camera system that captures and processes environmental data to support autonomous driving features while establishing protocols for employee access to customer recordings\\
         \hline
        \#14 & 23andMe Genetic Data Privacy & A direct-to-consumer genetic testing service that collects, analyzes, stores, and shares users' DNA data\\
         \hline
        \#15 & Starlink Satellite Surveillance & A satellite internet communication system for border security that enables continuous connectivity in remote areas, supports real-time data transmission from surveillance equipment, and facilitates agent communications while operating in regions with limited infrastructure\\
         \hline
        \#16 & Singapore TraceTogether & A nationwide contact tracing system that tracks citizens' movements and interpersonal contacts using smartphone applications and wearable tokens with data centrally stored in government databases\\
         \hline
        \#17 & Fog Reveal (Police) & A location data analytics platform that aggregates commercially available smartphone location information and provides law enforcement agencies with search capabilities to identify and track devices based on time, location, and movement patterns without requiring individual warrants\\
         \hline
        \#18 & Meta Facial Recognition & A social media photo tagging system that utilizes facial recognition technology to automatically identify individuals across a platform's user base and suggest tags based on facial geometry analysis from uploaded images and videos\\
         \hline
        \#19 & Ring Doorbell & A cloud-connected home security camera system that streams and stores video footage from users' homes, processes this data for motion detection alerts, and provides remote access capabilities through mobile applications with two-way audio communication features\\
         \hline
        \#20 & Reverse location search from photos & An AI-based image analysis system that can determine geographic location based on visual elements in photographs\\
         \hline
        \#21 & Dutch Child Care Fraudster Detection & An automated fraud detection system for government benefits programs that uses citizen data to create risk profiles and flag potentially fraudulent applications for investigation\\
         \hline
        \#22 & Telegram User Data Privacy Issues & A messaging platform that provides encrypted communications while managing legal compliance requirements regarding user data and content moderation\\
         \hline
        \#23 & League of Legends Chat-log Review & A system that analyzes employee behavior in company products during their personal time and incorporates this data into performance evaluations and employment decisions\\
         \hline
        \#24 & MiHoYo In-app Purchases & Mobile game with in-app purchases that appeals to users of all ages and maintains profitability\\
         \hline
        \#25 & Dark Patterns in Subscription Services & A subscription management system for digital services\\
         \hline
        \#26 & Olympic AI Surveillance & A video monitoring system that uses artificial intelligence to scan large crowds at major public events and automatically identify suspicious behaviors, unusual activities, and potential security threats\\
         \hline
        \#27 & Social Credit Score & A comprehensive monitoring and evaluation system for federal employees and contractors that tracks workplace performance, personal conduct, social media activity, and outside associations to generate trustworthiness scores influencing employment decisions\\
         \hline
        \#28 & Parental Control Applications & A parental control application for monitoring children's device usage\\
         \hline
        \#29 & Browser Fingerprint & A cross-site tracking system that identifies users without cookies by collecting technical information about their browsers and devices to create persistent identifiers that work even when privacy protections are enabled\\
         \hline
        \#30 & Government Email Transmission & An employee data management system for municipal governments that enables sharing workforce information with other government entities while managing sensitive personal details across organizational boundaries\\
    \bottomrule
    
    \end{tabular}
\label{tab:dataset}
\end{table*}

\clearpage
\section{Interview Questions for the Observational and Evaluation Study}
\label{sec:interview_questions}

\begin{table*}[htbp]
    \centering\small
\caption{The interview protocol used during the observational study.}
\Description{A table outlining the questions asked to participants during the observational study. It is organized by Phase / Section (e.g., Introduction, Post-study Questions, WS2: Assessing System Design) and lists the specific Prompts or questions for each part of the interview.
}
    \begin{tabular}{p{4cm} p{12cm}}
    \toprule
    \textbf{Phase / Section} & \textbf{Prompts} \\
    \midrule

    \textbf{Introduction} & Thank you for participating. In this study, we are interested in understanding how developers perform privacy risk assessments. We will ask you to work through a task using a standard framework called PRAM. There are no right or wrong answers; we are interested in your natural thought process. \\
    \midrule

    \textbf{Instruction} & As you work on the task, please try to think out loud. Tell us what you're thinking. This will help us understand your perspective. \\
    \midrule
    
    \textbf{Consent Script} & Before we begin the study, I need to obtain your informed consent. \newline
    This session will be recorded for the sole purpose of accurate transcription. The recording will be permanently deleted by the research team immediately after the transcript is verified. All data collected from the interview will be fully anonymized to protect your privacy. \newline
    Your participation is completely voluntary. You have the right to skip any question you are not comfortable with, or to withdraw from the study at any time without penalty. \\
    \midrule


    \multicolumn{2}{l}{\textbf{Post-study Questions}} \\
    \midrule

    Overall & 
    How would you describe your overall experience of using the PRAM framework? \newline
    Which of the PRAM worksheets or tasks did you find the most challenging, and why? \newline
    Which part of the framework, if any, did you find helpful or easy to use? Why? \newline
    If you could change one thing about this framework to make it easier for developers, what would it be? 
    \\
    \cmidrule(r){1-2}
    WS1: Framing Business Objectives & 
    How are you thinking about the main purpose of this system? \newline
    What factors are you considering when describing the functional requirements? \\
    \cmidrule(r){1-2}

    WS2: Assessing System Design & 
    As you map the data flows, could you explain: \newline
       \hspace*{2mm} - How did you decide which system components were important to include? \newline
    As you identify the data actions, could you explain: \newline
       \hspace*{2mm} - What was your process for identifying all the data actions here? \newline
       \hspace*{2mm} - How clear are the definitions for terms? \\
    \cmidrule(r){1-2}

    WS3: Prioritizing Risk & 
    As you assign Likelihood/Impact scores: \newline
       \hspace*{2mm} - Could you explain your reasoning for giving 'likelihood' that specific rating? \newline
    As you calculate the final risk score: \newline
       \hspace*{2mm} - What does that number tell you about the risk? \newline
       \hspace*{2mm} - How do you decide if this score is high enough to require action? \\
    \bottomrule
    \end{tabular}
    \label{tab:interview_protocol_s1}
\end{table*}

\begin{table*}[htbp]
    \centering\small
\caption{The interview questions used during the evaluation study.}
\Description{A table listing the post-study interview questions for the evaluation study comparing PRAM and \name. The columns are Theme (e.g., Overall Experiences, Strengths \& Weaknesses) and the Post-Study Interview Questions corresponding to each theme.}
    \begin{tabular}{p{3.5cm} p{12.5cm}}
    \toprule
    \textbf{Theme} & \textbf{Post-Study Interview Questions} \\
    \midrule
    Overall Experiences & How would you describe your overall experience using the PRAM worksheet versus \name? \newline
    Which approach did you find more intuitive or easier to use, and why?   
    \\
    \cmidrule(r){1-2}
    Strengths \& Weaknesses & What aspects of the PRAM worksheet did you find most helpful or effective?
    \newline
    What aspects of PRAM worksheet were confusing, frustrating, or difficult to use?
    \newline
    What challenges or limitations did you encounter with the PRAM worksheet?
    \newline
     What aspects of \name did you find most helpful or effective?
    \newline
    What aspects of \name were confusing, frustrating, or difficult to use?
    \newline
    What challenges or limitations did you encounter with \name?
    \\
    \cmidrule(r){1-2}
    Reflection \& Preference &     If you need to use one approach in a real project, which would you choose, and why? \newline
    How would you improve either approach based on your experience?
    \\
    \bottomrule
    \end{tabular}
    \label{tab:evaluation_interview}
\end{table*}

\newpage
\section{Qualitative Analysis Codebook from Observational Study}
\label{sec:codebook}

\begin{table*}[htbp]
    \centering\small
\caption{The codebook developed during our qualitative analysis of the observational study}
\Description{A table summarizing the themes and codes developed from analyzing user study data. The columns are Theme (e.g., Attention Allocation), Code (e.g., Information Overload), Sub-code (e.g., Cognitive Fatigue), and a Description of the observed challenge.
}
    \begin{tabular}{p{1.5cm} p{2.2cm} p{2.9cm} p{9.8cm}}
    \toprule
    \textbf{Theme} & \textbf{Code} & \textbf{Sub-code} & \textbf{Description} \\
    \midrule

    {\textbf{Attention Allocation}} & Information\newline Overload & Cognitive Fatigue & Participants experienced heavy mental load from juggling numerous factors simultaneously. This led to growing fatigue as the tasks progressed. \\
    \cmidrule(l){3-4} 
    & & Context-Switching\newline Overhead & Participants experienced additional cognitive load when transferring information across worksheets or repeatedly referring back and forth between different parts of the task. \\
    \cmidrule(l){2-4} 
    & Ineffective\newline Prioritization & Fixation on Familiar\newline Concepts & Participants focused excessively on design aspects that aligned with their technical background, while neglecting less familiar but critical privacy risks. \\
    \cmidrule(l){3-4}
    & & Fixation on Early Tasks & Participants spent a disproportionate amount of time and mental energy on the initial stages of a task, leading them to rush through or neglect later stages. \\
    \midrule

    {\textbf{Inaccessible Privacy Knowledge}} & Knowledge Blindspot & Unaware of Key Concepts & Participants overlooked important privacy issues because they did not know these aspects needed to be considered.\\
    \cmidrule(l){2-4}
    & Knowledge Recall Failure & Missed Known Concepts & Participants knew about a relevant privacy concept but forgot to apply or include it during the assessment due to a slip of the mind.
\\
    \cmidrule(l){3-4}
    & & Memory challenges & Participants struggled to carry insights and considerations from one part of the assessment to a later stage, often due to being overwhelmed. \\
    \midrule

    {\textbf{Insufficient Guidance}} & Unclear\newline Expectations & Ambiguous Granularity & Participants felt unsure about the required level of technical detail for their responses, whether to be high-level or dive into specific implementations. \\
    \cmidrule(l){3-4}
    & & Ambiguous Scope & Participants were unclear about how comprehensive their responses should be or what exactly to include. \\
    \cmidrule(l){2-4}
    & Vague Terms & Ambiguous Language & Key terms used in the framework were perceived as overly technical, unclear, or poorly defined, causing confusion for participants.\\
    \cmidrule(l){3-4}
    & & Inconsistent Framing & Concepts were defined or presented in inconsistent ways in different parts of the framework. \\
    \midrule

    {\textbf{Uncertainty in Risk Evaluation}} & Difficulty\newline Assigning Scores & Inconsistent Scoring & Without guidance on how to weigh different organizational impacts, participants assigned widely varying scores to the same privacy risk.\\\cmidrule(l){2-4}
    & Difficulty\newline Interpreting Scores & Arbitrary Risk Thresholds	 & Participants struggled to determine which calculated risk scores required action, as the framework lacked thresholds for what constitutes a high, medium, or low risk.\\
\bottomrule

    \end{tabular}
    \label{tab:codebook_study1}
\end{table*}

\end{document}